\documentclass[12pt]{article}
\usepackage{graphicx}

\newcommand\pubnumber{SLAC--PUB--13756}
\newcommand\pubdate{August, 2009}

\def\SLAC{SLAC, 
    Stanford University, Menlo Park, CA 94025 USA}
\def\doeack{\footnote{Work supported by the US Department of Energy,
                     contract DE--AC02--76SF00515.}}

\def\Title#1{\begin{center} {\Large #1 } \end{center}}
\def\Author#1{\begin{center}{ \sc #1} \end{center}}
\def\Address#1{\begin{center}{ \it #1} \end{center}}

\newcommand\pubblock{\rightline{\begin{tabular}{l} \pubnumber\\
         \pubdate \end{tabular}}}
\newenvironment{Abstract}{\begin{quotation} \begin{center}
                       ABSTRACT
     \end{center}\bigskip  }{\end{quotation}}

\def\Acknowledgements{\bigskip  \bigskip \begin{center} \begin{large}
             \bf ACKNOWLEDGEMENTS \end{large}\end{center}}
\def\beq{\begin{equation}}
\def\eeq#1{\label{#1}\end{equation}}
\def\eeqn{\end{equation}}

\newenvironment{Eqnarray}%
   {\arraycolsep 0.14em\begin{eqnarray}}{\end{eqnarray}}
\def\beqa{\begin{Eqnarray}}
\def\eeqa#1{\label{#1}\end{Eqnarray}}
\def\eeqan{\end{Eqnarray}}
\def\CR{\nonumber \\ }

\def\leqn#1{(\ref{#1})}

\let\bar=\overbar

\def\etal{{\it et al.}}

\def\bra#1{\left\langle{ #1} \right|}
\def\ket#1{\left| {#1} \right\rangle}

\def\lsim{\mathrel{\raise.3ex\hbox{$<$\kern-.75em\lower1ex\hbox{$\sim$}}}}
\def\gsim{\mathrel{\raise.3ex\hbox{$>$\kern-.75em\lower1ex\hbox{$\sim$}}}}

\def\M{{\cal M}}

\def\hc{{\mbox{\rm h.c.}}}
\def\tr{{\mbox{\rm tr}}}
\def\half{\frac{1}{2}}
\def\thalf{\frac{3}{2}}

\def\del{\partial}
\def\Dslash{\not{\hbox{\kern-4pt $D$}}}
\def\dslash{\not{\hbox{\kern-2pt $\del$}}}

\def\ee{e^+e^-}

\def\msb{{\bar{\scriptsize M \kern -1pt S}}}

\def\drb{{\bar{\scriptsize D \kern -1pt R}}}

\makeatletter
\def\section{\@startsection{section}{0}{\z@}{5.5ex plus .5ex minus
 1.5ex}{2.3ex plus .2ex}{\large\bf}}
\def\subsection{\@startsection{subsection}{1}{\z@}{3.5ex plus .5ex minus
 1.5ex}{1.3ex plus .2ex}{\normalsize\bf}}
\def\subsubsection{\@startsection{subsubsection}{2}{\z@}{-3.5ex plus
-1ex minus  -.2ex}{2.3ex plus .2ex}{\normalsize\sl}}

\renewcommand{\@makecaption}[2]{%
   \vskip 10pt
   \setbox\@tempboxa\hbox{\small #1: #2}
   \ifdim \wd\@tempboxa >\hsize     % IF longer than one line:
       \small #1: #2\par          %   THEN set as ordinary paragraph.
     \else                        %   ELSE  center.
       \hbox to\hsize{\hfil\box\@tempboxa\hfil}
   \fi}

 \def\citenum#1{{\def\@cite##1##2{##1}\cite{#1}}}
\newcount\@tempcntc
\def\@citex[#1]#2{\if@filesw\immediate\write\@auxout{\string\citation{#2}}\fi
  \@tempcnta\z@\@tempcntb\m@ne\def\@citea{}\@cite{\@for\@citeb:=#2\do
    {\@ifundefined
       {b@\@citeb}{\@citeo\@tempcntb\m@ne\@citea\def\@citea{,}{\bf ?}\@warning
       {Citation `\@citeb' on page \thepage \space undefined}}%
    {\setbox\z@\hbox{\global\@tempcntc0\csname b@\@citeb\endcsname\relax}%
     \ifnum\@tempcntc=\z@ \@citeo\@tempcntb\m@ne
       \@citea\def\@citea{,}\hbox{\csname b@\@citeb\endcsname}%
     \else
      \advance\@tempcntb\@ne
      \ifnum\@tempcntb=\@tempcntc
      \else\advance\@tempcntb\m@ne\@citeo
      \@tempcnta\@tempcntc\@tempcntb\@tempcntc\fi\fi}}\@citeo}{#1}}
\def\@citeo{\ifnum\@tempcnta>\@tempcntb\else\@citea\def\@citea{,}%
  \ifnum\@tempcnta=\@tempcntb\the\@tempcnta\else
  {\advance\@tempcnta\@ne\ifnum\@tempcnta=\@tempcntb \else\def\@citea{--}\fi
    \advance\@tempcnta\m@ne\the\@tempcnta\@citea\the\@tempcntb}\fi\fi}
\makeatother

\def\spa#1#2{\langle #1 #2 \rangle}
\def\spb#1#2{[ #1 #2 ]}
\def\apb#1#2#3{\langle #1 |  #2 | #3 ]}

\def\Op{{\cal O}}

\begin{document}
\begin{titlepage}
\pubblock

\vfill
\Title{ Spin-Dependent Antenna Splitting Functions}
\vfill
\Author{Andrew J. Larkoski and Michael E. Peskin\doeack}
\Address{\SLAC}
\vfill
\begin{Abstract}
We consider parton showers based on radiation from 
QCD dipoles or `antennae'.  These showers are built from
 $2\to 3$ parton splitting processes.  The question then arises of 
what functions replace the 
Altarelli-Parisi splitting functions in this approach.  We give a detailed
answer to this question, applicable to antenna showers in which partons carry 
definite helicity, and to both initial- and final-state emissions.
\end{Abstract}
\vfill
\begin{center} to appear in  Physical Review {\bf D} \end{center}
\vfill
\end{titlepage}
\def\thefootnote{\fnsymbol{footnote}}
\setcounter{footnote}{0}
\tableofcontents
\newpage
\setcounter{page}{1}

\section{Introduction}

In the studies that are now being done to prepare for physics at the LHC, 
many new approaches have been proposed to the old problem of generating 
parton showers.  The workhorse event generators 
 PYTHIA~\cite{PYTHIA} and HERWIG~\cite{HERWIG} generate parton
showers by successive radiations from individual partons.  The `splitting
functions' that define the radiation pattern are taken to be the 
kernels in the Altarelli-Parisi equation~\cite{AP,Dokshitzer}.  
This guarantees that the 
radiation pattern is correct in the region in which two partons become 
collinear.  Marchesini and Webber pointed out that it is also important
to include color interference between emissions from different 
partons~\cite{MW}.
In the workhorse generators, this is implemented by angular ordering 
of emissions.

The program ARIADNE, by Andersson, Gustafson, L\"onnblad, and 
Pettersson, took a different approach, 
implementing color coherence
by considering the QCD dipole to be the basic object that 
radiates a parton~\cite{colordipoles,ARIADNE}.  The
basic branching process in a parton shower is then a 
splitting in which two partons forming a color dipole radiate a third
parton.
  This approach has been 
taken up recently by a number of authors.  It is the basis for the 
VINCIA shower by Giele, Kosower, and Skands~\cite{VINCIA} and the parton
shower implementation in SHERPA by Krauss and Winter~\cite{SHERPAdipoles}.
We are also developing a parton shower based on this approach~\cite{hydra}.
In the years between ARIADNE and the newer works, the term `dipole' has been
applied in QCD to a different strategy based on $1\to 2$ splittings with 
recoil taken up by a third particle~\cite{CataniSeymour}.  To avoid confusion,
we will follow \cite{VINCIA} in calling the initial two-parton state an 
`antenna' and a branching process with $2\to 3$ splittings an `antenna 
shower'.

Central to the antenna shower is the $2\to 3$ splitting function, the function
that gives the relative branching probabilities as a function of the final
momenta.    The original ARIADNE program used an {\it ad hoc} proposal
satisfying the basic consistency requirements. 
It would be better to have a prescription that can be directly 
derived from QCD.
Splitting to three partons has been studied in great detail in the QCD
literature,
but not for this application.  Collinear systems of three partons are a 
part of the infrared structure of QCD at next-to-next-to-leading order,
and calculations that reach this level need an explicit prescription for
treating this set of infrared singularities.  Kosower~\cite{Kosower} 
defined the `antenna function' as a basic starting point for the 
analysis of this problem.  Many authors have computed 
antenna functions~\cite{Campbell,Catani,Duhr}.
Quite recently, Gehrmann-De Ridder, Gehrmann, and Glover have built a complete
formalism of `antenna subtraction' for NNLO calculations~\cite{Gehrmann}.  
The kernel in their theory can be 
interpreted as a $2\to 3$ splitting function, and it has been used 
to perform $2\to 3$ splitting in the VINCIA shower~\cite{VINCIA}.

In this paper, we will take a much more direct route to the construction
of $2\to 3$ splitting functions.  We will compute these functions by 
writing local operators that create two-parton final states and computing
their 3-parton matrix elements.  These calculations are very straightforward.
They can be used to treat individually all possible sets of polarized 
initial and final partons.  

This paper is organized as follows:  In Section 2, we will present our 
complete set of  spin-dependent
 $2\to 3$ splitting functions.   In Section 3,
we will give the derivation for the cases with total spin zero.  In 
Sections 4 and 5, we will give the derivation for the cases with nonzero total
 spin.

All of these derivations will be done in the kinematics of final-state
radiation.  This is the easiest situation to visualize and understand.
However, the same splitting functions can be used, after crossing, to describe 
parton emissions that involve initial-state particles.  We will explain 
how to use our expressions for initial-state showers in Section 6.

The $1\to 2$ Altarelli-Parisi 
splitting functions are universal in the sense that they result from a 
well-defined singular limit of QCD amplitudes.  For $2\to 3$ splitting 
functions there is no such universality. The collinear and soft limits
must agree with the known universal values, but away from these limits 
there is no unique prescription.  Earlier in this introduction, we 
made reference to a number of previous proposals for the spin-averaged
antenna splitting functions.  All of these, 
including the ARIADNE splitting functions, have 
the correct soft and collinear limits and so satisfy the basic requirements.
  In Section 7, we
will give a detailed comparison of the
 $2\to 3$ splitting functions obtained using our method
to  previous proposals for these splitting functions.

\section{Proposal for the $2\to 3$ splitting functions}

We begin by defining variables for $2\to 3$ splitting.  There are 
three cases of splittings that are needed for antenna showers: the 
final-final (FF) splitting, in which a third particle is created by 
coherent radiation from a two-particle system in the final state; 
the initial-final (IF) splitting, in which a third particle is created by 
coherent radiation from an initial- and a final-state particle; and
initial-initial (II) splitting, in which a third particle is created by 
coherent radiation from two initial-state particles.   It is easiest 
to understand the kinematics of antenna splitting for the FF case.
In this section, we will explain this kinematics and give a precise
prescription for the splitting functions.  In Section 6, we will extend
our prescription to the IF and II cases, in such a way that the same 
splitting functions can be used in those cases.

\begin{figure}
\begin{center}
\includegraphics[height=2.0in]{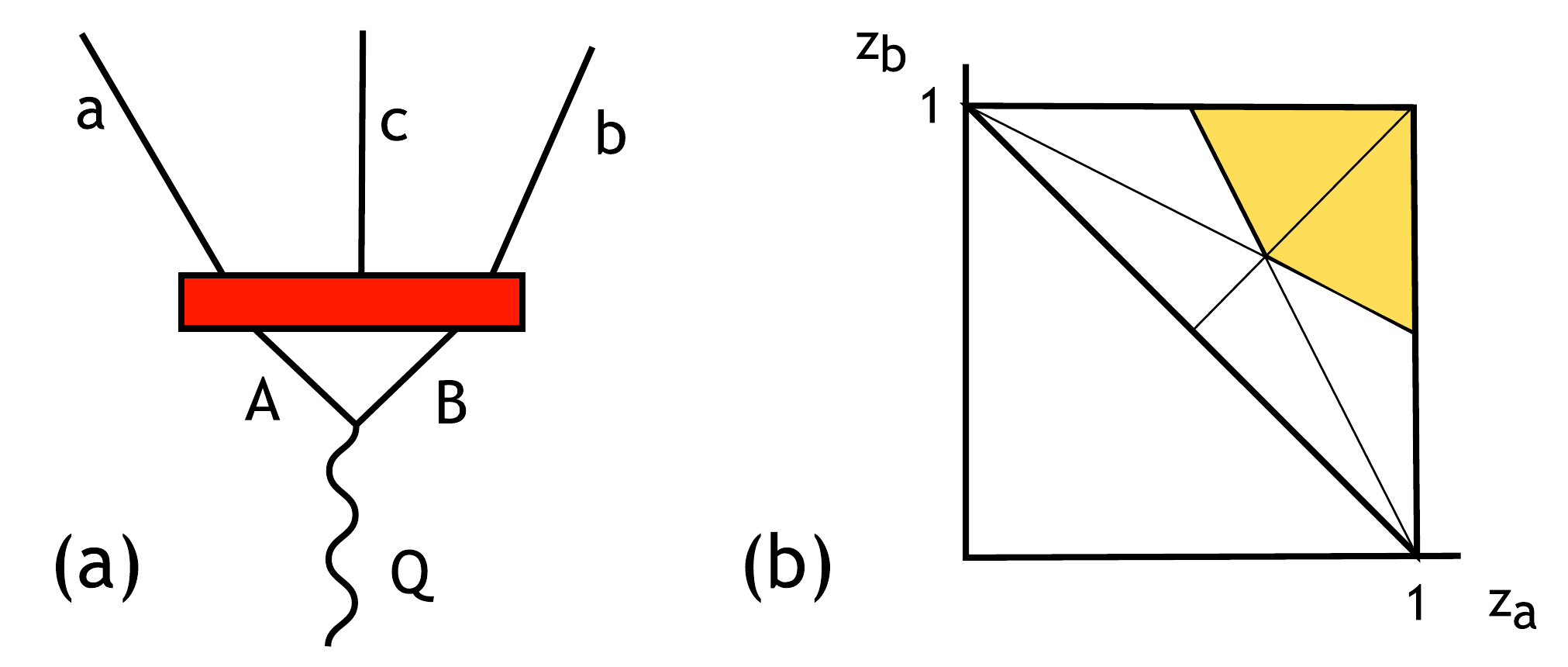}
\caption{(a) Kinematics of $2\to 3$ splitting in the final state (FF) case.
    (b) 
  Phase space for $2\to 3$ splitting in the FF case.  The 
        six regions corresponding to different orderings of 
        $s_{ab}$, $s_{ac}$, $s_{bc}$ are shown. The region that should
       be well described by an antenna splitting $AB\to acb$ is shaded.}
\label{fig:FFkinematics}
\end{center}
\end{figure}

Consider, then, a two-parton final-state system 
$(A,B)$ that splits to a 3-parton system $(a, c, b)$, conserving 
momentum, as shown in Fig.~\ref{fig:FFkinematics}(a). 
 Let $s_{ij} = (k_i + k_j)^2$, and let $Q = k_A + k_B = k_a + k_b + k_c$.

%%%%%%%%%%%%%%%%%%%%%%%%%%%%%%%%%%%%%%%%%%%%%%%%%%%%%%%%%%%%%%%%%%%%%%%%%
%%%%%%%%%%%%%%%%%%%%%%%%%%%%%%%%%%%%%%%%%%%%%%%%%%%%%%%%%%%%%%%%%%%%%%%%%%%

  The fractional invariant
masses in the final state are
\beq
    y_{ab} =  {s_{ab}\over s_{AB}} \ , \quad 
    y_{ac} =  {s_{ac}\over s_{AB}} \ , \quad 
    y_{bc} =  {s_{bc}\over s_{AB}} \ .
\eeq{ydefs}
The momentum fractions of the three particles in the $(AB)$ frame are
\beq
    z_{a} =  {2 Q \cdot k_a\over s_{AB}} \ , \quad 
    z_{b} =  {2 Q \cdot k_b\over s_{AB}} \ , \quad 
    z_{c} =  {2 Q\cdot k_c\over s_{AB}} \ .
\eeq{zdefs}
These obey the identities
\beq
   y_{ab} = (1 - z_c)\ ,  \quad  y_{ac} = (1 - z_b)\ , \quad
                      y_{bc} = (1 - z_a)\ .
\eeq{yzids}
and
\beq
   y_{ab} + y_{ac} + y_{bc} = 1 \ , \quad  z_a + z_b + z_c = 2 \ .
\eeq{yzmoreids}

The FF phase space covers the triangle $z_a \leq 1$, $z_b \leq 1$, 
     $z_a + z_b \geq 1$.  We can divide this phase space into 
     six triangles, each of which has a different ordering of the
         three quantities $y_{ab}$, $y_{ac}$, $y_{bc}$, as 
         shown in Fig.~\ref{fig:FFkinematics}(b).   An 
       antenna shower should give an accurate description 
     of the dynamics in  the two regions   $y_{ac} < y_{bc} < y_{ab}$, 
 $y_{bc} < y_{ac} < y_{ab}$ that are shaded in the figure. 
 
 A general problem in the generation of QCD radiation is that of 
possible double-counting.  Consider, for example, the process 
$\ee \to q g g \bar q$.  In some part of the phase space, the first
$g$ can be considered to be 
 radiated from the antenna of the $q$ and the second $g$; in 
another, the second $g$ can be considered to be radiated from the 
first $g$ and the $\bar q$.  These regions should be disjoint in the 
full 4-body phase space.  The complete solution to the problem is 
beyond the scope of this paper.  In simple terms, though, we can 
make the separation by choosing the radiated gluon to be softer
than the gluon from which it radiates.  This corresponds to 
integrating each antenna only over the shaded region
in  Fig.~\ref{fig:FFkinematics}(b).   A similar approximate
solution to the double-counting problem will apply 
in the other kinematic regions discussed in Section 6.   A more detailed
discussion of this issue can be found in \cite{VINCIA,hydra}.

Radiation from different QCD antenna is strictly independent 
and non-interfering
only in the limit of a large number of colors in QCD, $N_c \gg 1$.  Keeping
only terms leading in $N_c$ is known to be a good approximation to full 
QCD in many circumstances.  In particular, parton shower algorithms are 
correct only to leading order in $N_c$.  In this paper, we will explicitly
work only to the leading order for large $N_c$.

 In the limit of large $N_c$,
the rate for a $2\to 3$ splitting is given by  a formula of the form
\beq
    N_c  {\alpha_s\over 4\pi} \int dz_a dz_b \ \cdot  {\cal S}(z_a,z_b,z_c) 
\eeq{splitformal}
For example, in $\ee\to q_- g_+ \bar q_+$,
\beq
   {1\over \sigma_0} {d \sigma\over d z_a d z_b} =  N_c 
            {\alpha_s\over 4\pi} { z_a^2\over (1-z_a)
           (1-z_b) } \ ,
\eeq{exampleqgq}
where $(a,c,b)$ are the $(q,g,\bar q)$, respectively, $-$ and $+$ denote
left- and right-handed helicity, and
$\sigma_0$ is the cross section for 
$\ee\to q_- \bar q_+$~\cite{correctoneoverN}. 
Eq. \leqn{splitformal} will be our basic formula of reference. 
Using this notation, we  can write the various $2\to 3$ splitting functions
as 
\beq
    {\cal S} =   {{\cal N}(z_a,z_b,z_c) \over y_{ab} y_{ac} y_{bc}} \ ,
\eeq{Sproposal}
where the numerator is a simple function of the  $z_i$.  For example, for
the splitting $q_- \bar q_+ \to q_- g_+ \bar q_+$ given above, 
\beq
  {\cal N} =    y_{ab} z_a^2 =  (1 - z_c) z_a^2  \ . 
\eeq{Nexample}

In Table~\ref{tab:allN}, we give our proposal for
the numerator functions for all possible
cases of massless quark and gluon splittings.
The expressions are all monomials in the $y_{ij}$ and $z_j$. 

 In the FF kinematics, all
of the $y_{ij}$ and $z_i$ are positive and so   
${\cal S}(z_a,z_b,z_c)$ in \leqn{Sproposal},
is always positive,  In IF and II kinematics, some $y_{ij}$ and $z_i$
will become negative.  In most cases, the correct prescription is to
take ${\cal S}(z_a,z_b,z_c)$ to be the absolute value of the 
expression in  Table~\ref{tab:allN}.  However, there is a line within the
IF region where $z_a$ or $z_b$ crosses from positive to negative values. 
A few entries in the Table  change sign across this line.
We recommend that those entries be set to zero when $z_a$ or $z_b$ are 
negative.  We will give a detailed discussion of these points in 
Section 6.

The splitting functions ${\cal S}$ must give the correct universal
behavior in the soft and collinear limits. In the soft limit, $z_c \to 0$,
the numerators must go to 1 if the flavor and helicity of the final 
partons $a$ and $b$ match those of the initial partons $A$ and $B$;
otherwise, the numerators must go to 0.  It is easy to check that this 
test is satisfied.

%%%%%%%%%%%%%%%%%%%%%%%%%%%%%%%%%%%%%%%%%%%%%%%%%%%%%%%%%%%%%%%%%%%%%%%%

\begin{table}
\begin{center}
\begin{tabular}{c|cccccccc}
 & $+++$ &  $++-$ & $+-+$ & $-++$ & $--+$  & $-+-$ & $+--$ & $---$ \\ \hline
$ g_+ g_+ \to ggg$ &
  1   &   $  y_{ac}^4 $& $   y_{ab}^4$  & $ y_{bc}^4$ 
                                  &   0   &   0   &   0  & 0  \\
$ g_- g_+ \to ggg $&
  0   &    0  & $ y_{bc}^4$ & $z_a^4$ &
           $  z_b^4$  &$ y_{ac}^4$ & 0& 0 \\
$ g_+ g_+ \to \bar q q  g$ & 
  -   &    -  & $ y_{ab}^3 y_{bc}$  &  $y_{ab} y_{bc}^3$  
                & -       &   0   &   0  & -  \\
$ g_- g_+ \to \bar q q  g$ &
  -   &    -  &  $ y_{ab} y_{bc}^3 z_b^2 $ & $ z_a^2 z_b^2 y_{ab} y_{bc}$
                          & -   &   0   &   0  & -  \\
$ q_- \bar q_+ \to q g \bar q$ &
  -   &    -   &   - &  $ y_{ab} z_a^2 $   &  $ y_{ab} z_b^2 $ & - & -
         & - \\
$ q_- \bar q_- \to q g \bar q$ &
  -   &    -   &   - &   -    &   -  & $y_{ab}^3$ &
          - & $y_{ab}$ \\
$ q_- g_- \to qgg$ &
  -   &   - & - &  0 & $   y_{ac}^4$  & $ y_{ab}^3 z_b$ & - & $z_a$ \\ 
$ q_- g_+ \to qgg$ &
  -   &  - & - &  $  z_a^3 $  & $ y_{ab} z_b^3$ & $y_{ac}^4$ 
                &   - & 0 \\ 
$ q_- g_- \to q \bar q q$&
  -   &   - & - &  -  & $ y_{ab}  y_{ac}^3$  & $y_{ab}^2 y_{ac} z_b$ 
                  & - & -  \\ 
$ q_- g_+ \to q \bar q q$&
  -   &   - & - &  -  & $ z_a y_{ab} y_{ac} z_b^2  $  & $ z_a y_{ab} y_{ac}^3$
           & -  &   -  \\ 
\end{tabular}
\caption{Numerator functions ${\cal N}(z_a,z_b,z_c)$ for the 
 spin-dependent
$2\to 3$ splitting functions $AB \to acb$: $ {\cal S } =  
{\cal N}/(y_{ab} y_{ac} y_{bc})$.  Each line gives a choice of $AB$. 
The labels denote the polarization of the
three final particles with the radiated particle $c$ in the center: 
$(h_a, h_c, h_b)$.  The empty columns are forbidden by quark chiral 
symmetry. By the P and C invariance of QCD, the same expressions
apply after exchanging $- \leftrightarrow +$,  $q\leftrightarrow \bar q$,
or $ABacb \leftrightarrow BAbca$.} 
\label{tab:allN}
\end{center}
\end{table}
%%%%%%%%%%%%%%%%%%%%%%%%%%%%%%%%%%%%%%%%%%%%%%%%%%%%%%%%%%%%%%%%%%%%%%%%

In the collinear limits, we will insist that each antenna 
has the collinear behavior required in QCD.  One often hears the 
following statement about soft and collinear limits:
In dipole splitting ($1\to 2$ emission), each dipole has the
correct collinear behavior but the correct soft behavior is obtained
by combining neighboring 
dipoles.  In antenna splitting ($2\to 3$  emission), each
antenna has the correct soft limit but the correct collinear behavior
is obtained by combining neighboring antennae.  However, in the 
large $N_c$ limit, which we take to guide our intuition, different antennae
are independent radiators with different, non-interfering, colors flowing
in them.  From the viewpoint of this limit,
 each antenna, separately, must give both
the correct pattern of soft radiation and the correct pattern of
collinear radiation. 
This philosophy differs from that of the ARIADNE 
group~\cite{colordipoles,ARIADNE} and of \cite{SHERPAdipoles}.  We will 
discuss this point further when we compare with their results in 
Section 7. 
  
The collinear radiation from a given hard gluon is
then the sum of two contributions, one from each of the two 
antennae to which that hard gluon belongs.  In the large $N_c$ limit, 
these correspond to radiation from the color and anticolor lines of 
the gluon.  A single antenna, which has one of these contributions,
then has $\half$ of the standard collinear emission rate.  This factor
of $\half$ enters the check will we perform in a moment.  The 
factor comes entirely from bookkeeping and is independent of the
question of double-counting discussed briefly earlier in this Section. 

We now discuss the check of collinear limits.
Consider the limit in which $c$ becomes collinear with $a$. In
this limit, 
\beq
z_c \to z\ ,\qquad z_a \to (1-z)\ , \quad
         z_b \to 1\ , \quad  y_{ac}\to 0 \ . 
\eeq{collinear}
 The $2\to 3$ splitting function must reduce to
\beq
    {\cal S} \to   {1\over y_{ac}}  P(z) \ ,
\eeq
where $P(z)$ is the relevant spin-dependent
 Altarelli-Parisi splitting function.
These were presented in the original Altarelli-Parisi paper~\cite{AP} and
are reviewed in Table~\ref{tab:Altarelli}.  The functions are 
normalized as in \leqn{splitformal}, and as described in the 
previous paragraph: We take the large $N_c$
limit and divide by 2 where necessary to give the contribution from one
QCD antenna.  The denominator of 
\leqn{Sproposal} tends to $y_{ac} z(1-z)$  in this limit.  Then it is 
easy to check that the numerators match correctly in all cases.  The limit
in which $c$ becomes collinear with $b$ can be checked in the same way.
 
%%%%%%%%%%%%%%%%%%%%%%%%%%%%%%%%%%%%%%%%%%%%%%%%%%%%%%%%%%%%%%%%%%%%%%%%

\begin{table}
\begin{center}
\begin{tabular}{c | cccc}
 & $++$ &  $-+$ & $+-$ & $--$ \\ \hline
$ g_+ \to gg$\ : 
 & $ 1 / z(1-z)$   &   $(1-z)^3/ z$ & $z^3 / (1-z)$ & 0 \\ 
$ g_+ \to q \bar q$\ :  
 &  -   &   $ (1-z)^2$ & $ z^2 $ & - \\ 
$ q_- \to gq$\ : 
 &  -   & - &   $(1-z)^2/z$ & $1/z$ \\ 
$ q_- \to qg$\ :   
&   -   & $z^2/(1-z)$ & - &   $1/(1-z)$ \\ 
\end{tabular}
\caption{Spin-dependent Altarelli-Parisi splitting functions $P(z)$
for splittings $B \to cb$.  The 
labels denote the polarization of the two final particles with the 
radiated particle first: $(h_c,h_b)$.  The empty columns are forbidden 
by quark chiral 
symmetry. By the P and C invariance of QCD, 
the same expressions
apply after exchanging $- \leftrightarrow +$ or $q\leftrightarrow \bar q$.}
\label{tab:Altarelli}
\end{center}
\end{table}
%%%%%%%%%%%%%%%%%%%%%%%%%%%%%%%%%%%%%%%%%%%%%%%%%%%%%%%%%%%%%%%%%%%%%%%%

When the collinear limits and the soft limit are all
 nonzero, there is a unique monomial
of the $y$'s and $z$'s that gives all limits correctly.  In the other
cases, there is some ambiguity.  In all cases, it would be desirable
if the results in Table~\ref{tab:allN} could be derived directly by 
simple Feynman diagram computations.  In the next few sections, we will
present those derivations.

\section{Spin-0 case}

To compute the $2\to 3$ splitting functions, we will use the following method:
Write an operator that, at the leading order, creates a 2-parton state 
with definite helicity.  Then, compute the 3-particle matrix element.
This realizes in a very simple way the splitting process illustrated in 
Fig.~\ref{fig:FFkinematics}.

To create massless quarks and antiquarks of definite helicity, we will use 
the appropriate chiral fermion fields.  To create gluons of definite 
helicity, we will use the operators
\beq
    \sigma \cdot F =  \half \sigma^m \bar  \sigma^n F_{mn} \ , \qquad
    \bar \sigma \cdot F =  \half \bar \sigma^m \sigma^n F_{mn} \ , 
\eeq{sigmaFs}
where $\sigma^m$, $\bar \sigma^m$ are the $2\times 2$ matrix entries of the
Dirac matrices in a chiral basis and $F_{mn}$ is the gluon field strength
tensor.  At leading order, $\sigma \cdot F$ creates a $+$ helicity gluon, 
and $\bar \sigma \cdot F$ creates a $-$ helicity gluon.

The 2-parton state $g_+g_+$ in the first line of Table~\ref{tab:allN} can
be created from the spin-0 operator
\beq 
           \Op =        \half  \tr [ (\sigma\cdot F)^2 ]  \ .
\eeq{firstop}
We can then compute the splitting function for this polarized initial state
explicitly from the definition
\beq
      {\cal S}(z_a, z_c, z_b) =   Q^2 \left| {\M(\Op \to acb)
              \over \M(\Op \to AB) } \right|^2  
\eeq{Sdefin}
In the next few sections, we will compute all of the splitting functions
in Table~\ref{tab:allN} using this formula, with a different choice of 
the operator $\Op$ for each line of the table.

To evaluate \leqn{Sdefin}, we need to compute the matrix elements
of $\Op$, with total momentum $Q$ injected, to  3-gluon final states.
The result can be expressed in terms of color-ordered amplitudes.  We 
identify the color-ordered amplitude that multiplies the color structure
$\tr [T^a T^c T^b]$ with the splitting function.  To carry out these 
computations, we will use the spinor product formalism.  That is, 
instead of working with 4-vectors, we will use as our basic objects
the spinor products
\beq 
\spa ij = \bar u_-(i) u_+(j)  \ , \qquad   \spb ij = \bar u_+(i) u_-(j)\ . 
\eeq{spinorproducts}
These objects obey 
\beq
|\spa ij|^2 = |\spb ij|^2 = s_{ij} \ .
\eeq{spsquare}
Methods for QCD
computations with spinor products and color-ordering are explained in 
\cite{ManganoParke,Dixon}.   In this notation, 
the matrix element for $\Op$ to create a 
$g_+g_+$ final state is 
\beq 
   \bra{g_+g_+}\Op \ket{0}  = {\spb AB}^2  \ .
\eeq{firstOnorm}

%%%%%%%%%%%%%%%%%%%%%%%%%%%%%%%%%%%%%%%%%%%%%%%%%%%%%%%%%%%%%%%%%%%%%%%%%
\begin{figure}
\begin{center}
\includegraphics[height=1.2in]{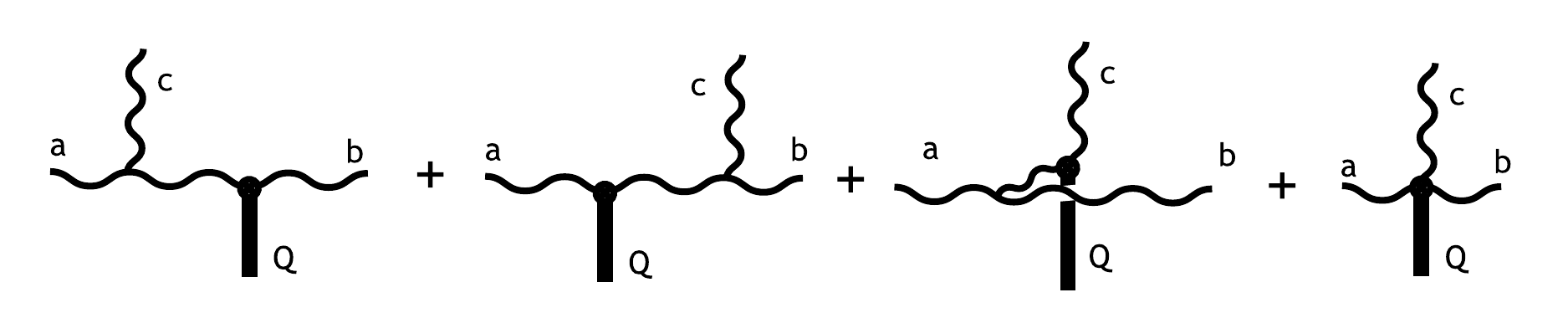}
\caption{Feynman diagrams for the computation of the $gg\to ggg$ splitting
 functions.}
\label{fig:ggg}
\end{center}
\end{figure}
%%%%%%%%%%%%%%%%%%%%%%%%%%%%%%%%%%%%%%%%%%%%%%%%%%%%%%%%%%%%%%%%%%%%%%%%%%%

The three-gluon matrix elements of the operator \leqn{firstop} are given
by the diagrams in Fig.~\ref{fig:ggg}.  These diagrams have already been 
analyzed by Dixon, Glover, and Khoze as a part of their analysis of 
the coupling of the Higgs boson to multi-gluon states~\cite{LanceGlover}.
They find
\beqa
  {\cal A}(\Op \to g_+g_+g_+)   &= &  {s_{AB}^2\over \spa ac \spa cb \spa ba}
               \CR
  {\cal A}(\Op \to g_+g_+g_-)   &= &
 {{ \spb ac}^4\over \spb ac \spb cb \spb ba}
               \CR
  {\cal A}(\Op \to g_+g_-g_+)   &= &
 {{ \spb ab}^4\over \spb ac \spb cb \spb ba}
               \CR
  {\cal A}(\Op \to g_-g_+g_+)   &= &
 {{ \spb bc}^4\over \spb ac \spb cb \spb ba}
               \CR
\eeqa{Aforggtogggzero}
and zero for the other four cases.  After squaring, using \leqn{spsquare},
and dividing by the square of \leqn{firstOnorm}, we obtain the first line of 
Table~\ref{tab:allN}.

One of the major points of  \cite{LanceGlover} is that the results
\leqn{Aforggtogggzero} belong to series of Maximally Helicity Violating
(MHV) amplitudes that have a simple form for any number of gluons emitted.
Actually, all of the amplitudes that we will compute in this paper are 
similarly simple and belong to MHV series.   The use of MHV amplitudes
to study antenna splitting is explored for higher-order processes
 in \cite{Duhr}. 

In principle, the initial state $g_+g_+$ could also have been created by 
an operator of spin 2, or some higher spin.  This would have led to a more
complicated expression for the $2\to 3$ splitting function, with, however,
the same soft and collinear limits.  This illustrates the ambiguity in the
definitions of $2\to 3$ splitting functions refered to in the introduction.
The simplest results are obtained using the operator of minimal spin, and 
we will make that choice in all of the examples to follow.

\begin{figure}
\begin{center}
\includegraphics[height=1.0in]{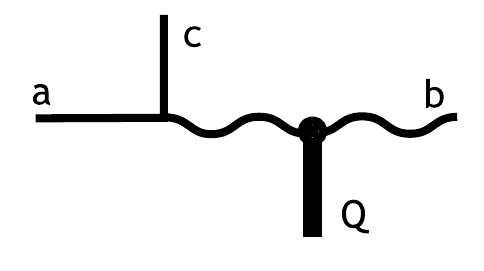}
\caption{Feynman diagram for the computation of the $gg \to \bar q q g$ 
splitting functions.}
\label{fig:qqg}
\end{center}
\end{figure}

The diagram shown in Fig.~\ref{fig:qqg} gives the splitting of the 
two-gluon initial state to $\bar q q g$.  We find
\beqa
  {\cal A}(\Op \to \bar q_+ q_- g_+)   &= & { {\spb ab}^2\over \spb ac }
               \CR
  {\cal A}(\Op \to  \bar q_- q_+ g_+)  &= &  { {\spb cb}^2\over \spb ac }
\eeqa{Aforggtoqqgzero}
There is no splitting to a final $g_-$.  This gives the result in the 
third line of the table. 

%%%%%%%%%%%%%%%%%%%%%%%%%%%%%%%%%%%%%%%%%%%%%%%%%%%%%%%%%%%%%%%%%%%%%%%%%
%%%%%%%%%%%%%%%%%%%%%%%%%%%%%%%%%%%%%%%%%%%%%%%%%%%%%%%%%%%%%%%%%%%%%%%%%%%

The initial state $q_- \bar q_- $ can also be created by a spin 0 operator
\beq
              \Op =      \bar q_L q_R \ .
\eeq{qqspinzero}
The matrix element for this operator to create a $q_- \bar q_-$ final state
is 
\beq 
   \bra{q_-\bar q_-}\Op \ket{0}  = \spa AB  \ .
\eeq{secondOnorm}
A straightforward calculation gives
\beqa
  {\cal A}(\Op \to q_- g_+ \bar q_-)   &= &
  { {\spa ab}^2\over \spa ac \spa cb } 
               \CR
  {\cal A}(\Op \to q_- g_- \bar q_-)   &= &  { s_{AB}\over \spb ac \spb cb } 
\eeqa{Aforqqtoqgqzero} 
These give the results shown in the sixth line of the table.

\section{Spin-1 and spin-2 case}

In \cite{colordipoles}, the $2\to 3$ splitting function for $q\bar q \to 
q g \bar q$ was derived from the cross section for $\ee\to q g \bar q$.
From the point of view of the previous section, this corresponds to creating
the 2- and 3-parton final states using the operator
\beq
       \Op  = \bar q_L \gamma^m q_L \ .
\eeq{spinoneoperator}
To obtain a definite matrix element, we must contract this operator with a
polarization vector.  A convenient choice is to introduce two new massless
vectors 1 and 2, such that $k_1 + k_2 = k_A + k_B$, and to choose the 
polarization vector to be  $\epsilon^\mu = \apb 1 {\gamma^\mu} 2 $.
This is effectively the procedure of decaying the massive vector that
couples to the operator \leqn{spinoneoperator} into a pair of massless
vectors to facilitate the analysis; this is a standard method in spinor
product calculations~\cite{KleissStirling}.  We then recast
\beq
       \Op  = \half \bar q_L \gamma^m q_L \  \apb 1 {\gamma_m} 2   \ .
\eeq{spinonetwo}

The matrix element of \leqn{spinonetwo} to a $q_- \bar q_+$ state is
\beq 
   \bra{q_-\bar q_+}\Op \ket{0}  = - \spa 1A \spb 2B  \ .
\eeq{thirdOnorm}
The direction of the 1-2 system
 chooses the helicity of the final partons.  In this
case, there is only one choice, and so the amplitude vanishes when $1$ is 
parallel to $A$ or $2$ is parallel to $B$.  This will not always be 
true in our later examples.  But, we will always be able to choose
the desired helicity of $A$ and $B$ by choosing
$1$ parallel to $B$  and $2$ parallel to $A$.
    
The matrix elements for the operator \leqn{spinonetwo}
to create  3-parton final states are
\beqa
  {\cal A}(\Op \to q_- g_+ \bar q_+)   &= &  { {\spa 1a}^2\spb 12
               \over \spa ac \spa cb } 
               \CR
  {\cal A}(\Op \to q_- g_- \bar q_+)   &= &  { {\spb 2b}^2 \spa 12 
        \over \spb ac \spb cb } \ .
\eeqa{Aforqqtoqgqone} 

To compute the results in the fifth line of the table, we must essentially
divide \leqn{Aforqqtoqgqone} by \leqn{thirdOnorm} and square the result.
To do this, we need a prescription for treating the expressions $\spa 1a$
and $\spb 2b$ in the numerators. The problem of relating 
the vectors $a$, $b$, $c$ to $A$ and $B$ in an antenna splitting was
discussed at length by Kosower in \cite{KosowerII}; that paper 
gives a general treatment in terms of {\it reconstruction functions}
to provide expressions that can be smoothly  integrated in higher-order
QCD calculations.  This discussion is generalized to 
the initial-state channels in \cite{Daleo}. 
Here, we will take a more {\it ad hoc} approach
that leads to the simplest formulae with correct singular limits.

Formulae for  $\spa 1a$
and $\spb 2b$ that are  simple and become
exact in the collinear and soft limits are found by  approximating
 $a$ collinear with
$A$ and $b$ collinear with $B$.  Then identifying $1$ with $B$ and $2$ with 
$A$ gives
\beq 
    |\spa 1a|^2  = s_{Ba} \to  z_a s_{AB}\ , \quad
    |\spa 1b|^2  \to 0 \ , \quad   |\spa 2a|^2  \to 0 \ , 
  |\spa 2b|^2  = s_{Ab} \to  z_b s_{AB}\ ,
\eeq{onetworeduce}
and similarly for the conjugate products.
Using this prescription, one obtains the fifth line of the table.  This is
a more formal version of the argument for these entries already given 
in Section~2.

In our calculations, we will encounter two more numerator objects that
 require reconstruction, namely,   $\spa 1c$ and $\spa 2c$.  
The prescription
above gives
\beq 
    |\spa 1c|^2  = s_{Bc} \to  (y_{bc}/z_b) s_{AB}\ , \quad
    |\spa 2c|^2  = s_{Ac} \to  (y_{ac}/z_a) s_{AB}\ .
\eeq{onetworeducetwo}
However, it is potentially dangerous to write factors of $z_a$, $z_b$ in the 
denominator.  We will see in Section 6 that such factors would create
unphysical singularities when continued to the IF kinematics. 
These unphysical singularities are avoided in the general formalism 
used in \cite{KosowerII}, but at the price of introducing much more 
complicated formulae.   Fortunately, 
we will see that $\spa 1c$ arises only in situations where there is no 
collinear singularity with $c$ parallel to $b$.  In such cases, the remaining
universal singular terms---the collinear singularity with $c$ parallel to 
$a$ and the soft singularity---correspond to kinematic limits with $z_b \to 1$.
A similar consideration applies to $\spa 2c$.  Thus, we choose, instead 
of using \leqn{onetworeducetwo}, to evaluate these quantities as
\beq 
    |\spa 1c|^2  = s_{Bc} \to  y_{bc} s_{AB}\ , \quad
    |\spa 2c|^2  = s_{Ac} \to  y_{ac} s_{AB}\ .
\eeq{onetworeducethree}
This gives an incorrect shape in a  region where $a$  and $b$ are 
collinear, but, hopefully, we will not use the $AB \to acb$ splitting function
to evaluate the rate to fill this region of phase space.

Another choice for evaluating $\spa 1c$ and $\spa 2c$ is to
replace both  expressions by $z_c$. 
However, the spinor product $\spa 1c$  vanishes in the $bc$ collinear
limit but not in the $ac$ collinear limit, and conversely for
$\spa 2c$, so this choice does not give the universal singularities 
correctly.

We now apply this formalism to compute  the second and fourth lines
of Table~\ref{tab:allN}, associated with the $g_-g_+$ antenna.  This 
antenna is created by the spin-2 operator  $\tr [ \gamma^m (\bar\sigma \cdot F)
 \gamma^n ( \sigma \cdot F) ] $.  To make a definite calculation, we 
need a spin-2 polarization vector.  An appropriate choice can be found by 
introducing the massless vectors 1 and 2 as above and writing
\beq
    \epsilon^{mn}  =  \apb 1 {\gamma^m} 2 \  \apb 1 {\gamma^n} 2   \ .
\eeq{spintwoepsilon}
This effectively decays the masive spin-2 particle into two massless spinors.
This
 method was introduced in \cite{LePeskin} to compute the relevant amplitudes
for the emission of massive gravitons at high-energy colliders.

With this prescription, we generate the $g_-g_+$ antenna using the operator
\beq
       \Op  = {1\over 4}\tr [ \gamma^m (\bar\sigma \cdot F)
 \gamma^n (\sigma \cdot F) ]  \apb 1 {\gamma_m} 2   \apb 1 {\gamma_n} 2 
\eeq{spintwo}
The matrix element of this operator that creates the 2-parton dipole is
\beq 
   \bra{g_- g_+}\Op \ket{0}  = {\spa 1A}^2 {\spb 2B}^2  \ .
\eeq{fourthOnorm}
To obtain the correct initial polarizations, we take $1 = B$, $2 = A$ as
before. 
 The matrix elements to the possible 3-parton final states are
\beqa
  {\cal A}(\Op \to g_+g_+g_+)   &= &  0  \CR
  {\cal A}(\Op \to g_+g_+g_-)   &= &  { {\spa 1b}^4{\spb 12}^2
               \over \spa ab \spa ac \spa cb }  \CR
  {\cal A}(\Op \to g_+g_-g_+)   &= &  { {\spa 1c}^4{\spb 12}^2
               \over \spa ab \spa ac \spa cb }  \CR
  {\cal A}(\Op \to g_-g_+g_+)   &= &  { {\spa 1a}^4{\spb 12}^2
               \over \spa ab \spa ac \spa cb } \ ,
\eeqa{Aforgggtwo} 
and the conjugates with $1 \leftrightarrow 2$ for the other four combinations.
Applying the reductions \leqn{onetworeduce}, \leqn{onetworeducetwo}, we find
the results given in the second line of the table.

The nonzero matrix elements of this operator to $\bar q q g$ final states are
\beqa
  {\cal A}(\Op \to \bar q_+ q_- g_+)&=&   { {\spa 1c}^2{\spb 2b}^2
               \over \spb ac }   \CR
  {\cal A}(\Op \to \bar q_- q_+ g_+) &=& {{\spa 1a}^2 {\spb 2b}^2
               \over \spb ac }  \ .
\eeqa{Aforqqgtwo} 
The same reduction process gives the results in the fourth line of the
table.

\section{Spin-$\half$ and spin-$\thalf$ cases}

%%%%%%%%%%%%%%%%%%%%%%%%%%%%%%%%%%%%%%%%%%%%%%%%%%%%%%%%%%%%%%%%%%%%%%%%%
\begin{figure}
\begin{center}
\includegraphics[height=1.2in]{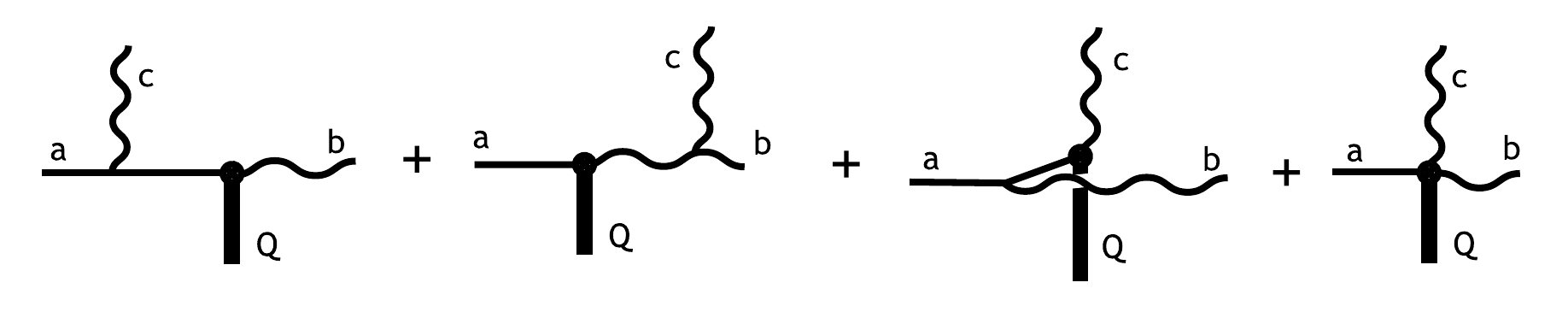}
\caption{Feynman diagrams for the computation of the $qg\to qgg$ splitting
 functions.}
\label{fig:qgg}
\end{center}
\end{figure}
%%%%%%%%%%%%%%%%%%%%%%%%%%%%%%%%%%%%%%%%%%%%%%%%%%%%%%%%%%%%%%%%%%%%%%%%%%%

The cases of quark-gluon antennae can be treated in the same way.  There is
one additional subtlety.  In QCD, quarks  are color triplets and gluons 
are color octets, so a quark-gluon operator carries net color.  This 
means that the matrix element for gluon emission from a 
quark-gluon operator is not gauge-invariant unless we allow the gluon also
to be emitted from the initial state.  This makes it unclear how to 
define a quark-gluon antenna.

We resolve this problem with the following prescription:  We consider the
quarks to be color octet particles like the gluons.  Then, as in the 
previous sections, we extract the color-ordered contribution corresponding 
to emission from the antenna.  In the limit of large $N_c$, the various
antennae in a process radiate independently.  The diagrams contributing
to a quark-gluon antenna in this prescription are shown in Fig.~\ref{fig:qgg}.
The third diagram, with an intermediate quark line, does not appear 
in QCD.  However, it does nicely provide the missing piece that makes this
sum of diagrams gauge-invariant without radiation from the initial state.

This solution is the same as that found in the earlier work of 
Gehrmann-De Ridder, Gehrmann, and Glover~\cite{Gehrmann}.
  Those  authors computed the 
quark-gluon antennae by factorizing the amplitudes for
the decay of a neutralino into a massless gluino plus $gg$ or $q\bar q$.
In their calculation, the off-shell color octet fermion is the 
gluino. 

With this understanding, we proceed as in the previous Section. 
We can generate the $q_-g_-$ antenna using the operator $\bar q_L 
(\bar\sigma\cdot
F)$.  The polarization spinor can be built by introducing massless
spinors 1 and 2 as above and taking $\ket{2}$ to be this spinor. Then
\beq
       \Op  = -i \bar q_L (\bar \sigma \cdot F) \ket{2} \ .
\eeq{spinhalf}
The matrix element of this operator that creates the 2-parton dipole is
\beq 
   \bra{q_- g_-}\Op \ket{0}  = {\spa AB} {\spb B2}  \ .
\eeq{fifthOnorm}
To obtain the correct initial polarizations, we take $1 = B$, $2 = A$.

 The matrix elements to the possible 3-parton final states are
\beqa
  {\cal A}(\Op \to q_-g_+g_+)   &= &  0  \CR
  {\cal A}(\Op \to q_- g_-g_+)   &= &  { {\spa ac}^3{\spa 2c}
               \over \spa ab \spa ac \spa cb }  \CR
  {\cal A}(\Op \to q_-g_+g_-)   &= &  { {\spa ab}^3{\spa 2b}
               \over \spa ab \spa ac \spa cb }  \CR
  {\cal A}(\Op \to q_-g_-g_-)   &= &  { s_{AB} \spa 12 \spb 1a 
               \over \spb ab \spb ac \spb cb } \ .
\eeqa{Aforqgghalf} 
Applying the reductions \leqn{onetworeduce}, \leqn{onetworeducetwo}, we find
the results given in the seventh line of the table.

The nonzero matrix elements of this operator to $ q \bar q q$ final 
states are
\beqa
  {\cal A}(\Op \to q_- \bar q_- q_+)&=&   {{\spa ac} {\spa 2c}
               \over \spa cb }  \CR
  {\cal A}(\Op \to q_- \bar q_+ q_-) &=& - { {\spa ab}{\spa 2b}
               \over \spa cb }  \ .
\eeqa{Aforqqqhalf} 
The same reduction process gives the results in the ninth line of the
table.

We generate the $q_-g_+$ antenna using the spin-$\thalf$
operator $\bar q_L \gamma^m (\sigma\cdot
F)$.  This is essentially the supersymmetry current of the system of 
gluons and color octet fermions.
  The polarization spinor can be built by introducing massless
spinors 1 and 2 as above:
\beq
       \Op  = i \bar q_L \gamma^m (\sigma \cdot F) \, 2] \apb 1{ \gamma_m}2
         \ .
\eeq{spinthalf}
The matrix element of this operator that creates the 2-parton dipole is
\beq 
   \bra{q_- g_+}\Op \ket{0}  = {\spa 1A} {\spb 2B}^2  \ .
\eeq{sixthOnorm}
To obtain the correct initial polarizations, we again take $1 = B$, $2 = A$.

 The matrix elements to the possible 3-parton final states are
\beqa
  {\cal A}(\Op \to q_-g_+g_+)   &= &  { {\spa 1a}^3{\spb 12}^2
               \over \spa ab \spa ac \spa cb }  \CR 
  {\cal A}(\Op \to q_- g_-g_+)   &= &  { \spa ab {\spb 2b}^3{\spa 12}
               \over \spb ab \spb ac \spb cb }  \CR
  {\cal A}(\Op \to q_-g_+g_-)   &= &  { \spa ac {\spb 2c}^3{\spa 12}
               \over \spb ab \spb ac \spb cb }  \CR
  {\cal A}(\Op \to q_-g_-g_-)   &= &  0 \  . 
\eeqa{Aforqggthalf} 
Applying the reductions \leqn{onetworeduce}, \leqn{onetworeducetwo}, we find
the results given in the eighth line of the table.

The nonzero matrix elements of this operator to $ q \bar q q$ final 
states are
\beqa
  {\cal A}(\Op \to q_- \bar q_- q_+)&=&   {{\spa 1a} {\spb 2b}^2
               \over \spb cb }  \CR
  {\cal A}(\Op \to q_- \bar q_+ q_-) &=& - { {\spa 1a} {\spb 2c}^2
               \over \spb cb }  \ .
\eeqa{Aforqqqthalf} 
The same reduction process gives the results in the tenth line of the
table.

\section{Initial-state showers}

The Feynman diagram computations that we have done to find the antenna
splitting functions for FF splittings can also be applied, by crossing,
to IF and II splittings.  The expressions in Table~\ref{tab:allN} are 
given in terms of invariant quantities that are unchanged under
crossing.  Thus, we can use the expressions in this table directly in 
other channels.  At worst, a change of the overall sign is required in 
some cases.  In this section, we will clarify this statement by analyzing
the kinematics of IF and II splittings in the same variables as those used
in Section 2 for FF splittings. In all cases, the kinematics is done for 
all massless partons only.   The kinematic discussion in this section 
is similar to that presented in \cite{Daleo}.

To begin, we will formalize some of the results quoted in Section 2
for the FF region.  The cross section for a process $X \to acb$ is
\beq
   \sigma(X\to acb) = {1\over \Phi_X} {s\over 128 \pi^3} \int dz_a dz_b
        | \M(X\to acb)|^2 \ ,
\eeq{basicFF}
where $\Phi_X$ is the flux factor. Polarization and color indices have
been suppressed.  The left-hand side has been integrated over the 
orientation of the final state system but is otherwise exact.  To 
write an expression involving the antenna splitting function, we 
approximate
\beq
     \M(X\to acb) \approx    \M(X \to AB) \cdot g T \cdot {\M(\Op \to acb)
              \over \M(\Op \to AB) } \ , 
\eeq{MdecompFF}
where  $\Op$ is the operator used in Sections 3--5 to represent the state
$AB$. The factor $gT$ is the QCD coupling and color matrix; after squaring
and summing over colors, this becomes $4\pi \alpha_s N_c$.  The splitting
function is  defined by \leqn{Sdefin}, 
\beq
            {\cal S}(z_a, z_c, z_b) =   s_{AB} \left| {\M(\Op \to acb)
              \over \M(\Op \to AB) } \right|^2 
\eeq{Sdefintwo}
Then 
\beq
   \sigma(X\to acb) \approx \sigma(X\to AB)\cdot
  {\alpha_s N_c\over 4\pi}\int dz_a dz_b  {\cal S}(z_a, z_c, z_b) \ .
\eeq{finalFF}
It is important to note that, in this formula or in \leqn{MdecompFF}, 
the vectors $k_A$ and
$k_B$ are introduced as part of the approximation. They can be defined
in any way that is consistent with the requirements that $k_A$ and $k_B$
are lightlike, $k_A+k_B = Q$, and $k_A$ and $k_B$ become parallel to 
$k_a$ and $k_b$, respectively, in the soft and collinear limits. 

The logic of this derivation extends straightforwardly to the IF and II 
regions. The major change is that, in  these cases, we 
need to introduce initial hadrons from which 
the initial partons are extracted.

Consider first the IF case.  The cross section for a proton of momentum 
$P$ to scatter from 
a color-singlet system $X$ transferring momentum $Q$ to create a 2-parton
system $cb$ is
\beq
\sigma (p X \to cb) = \int dx_a  f(x_a)\  {1\over \Phi_{aX}} {1\over 16\pi}
\int d \cos\theta_* \  | \M(aX\to cb)|^2 \ ,
\eeq{basicIF}
where $\cos\theta_*$ is the scattering angle in the $cb$ center of mass
system.
We will approximate this formula using the expression
analogous to \leqn{MdecompFF}
\beq
     \M(aX\to cb) \approx    \M(AX \to B) \cdot g T \cdot {\M(a\Op \to cb)
              \over \M(A\Op \to B) } \ . 
\eeq{MdecompIF}
Then the splitting function  is defined by 
the same expression ${\cal S}$ as in  \leqn{Sdefintwo}, but now 
analytically continued into the new kinematic region.  If a fermion line
is crossed from the final to the initial state, an extra factor (-1) should
be included.  In addition, $s_{AB}$ in  \leqn{Sdefintwo} is negative
in this region, giving an extra minus sign.

\begin{figure}
\begin{center}
\includegraphics[height=2.0in]{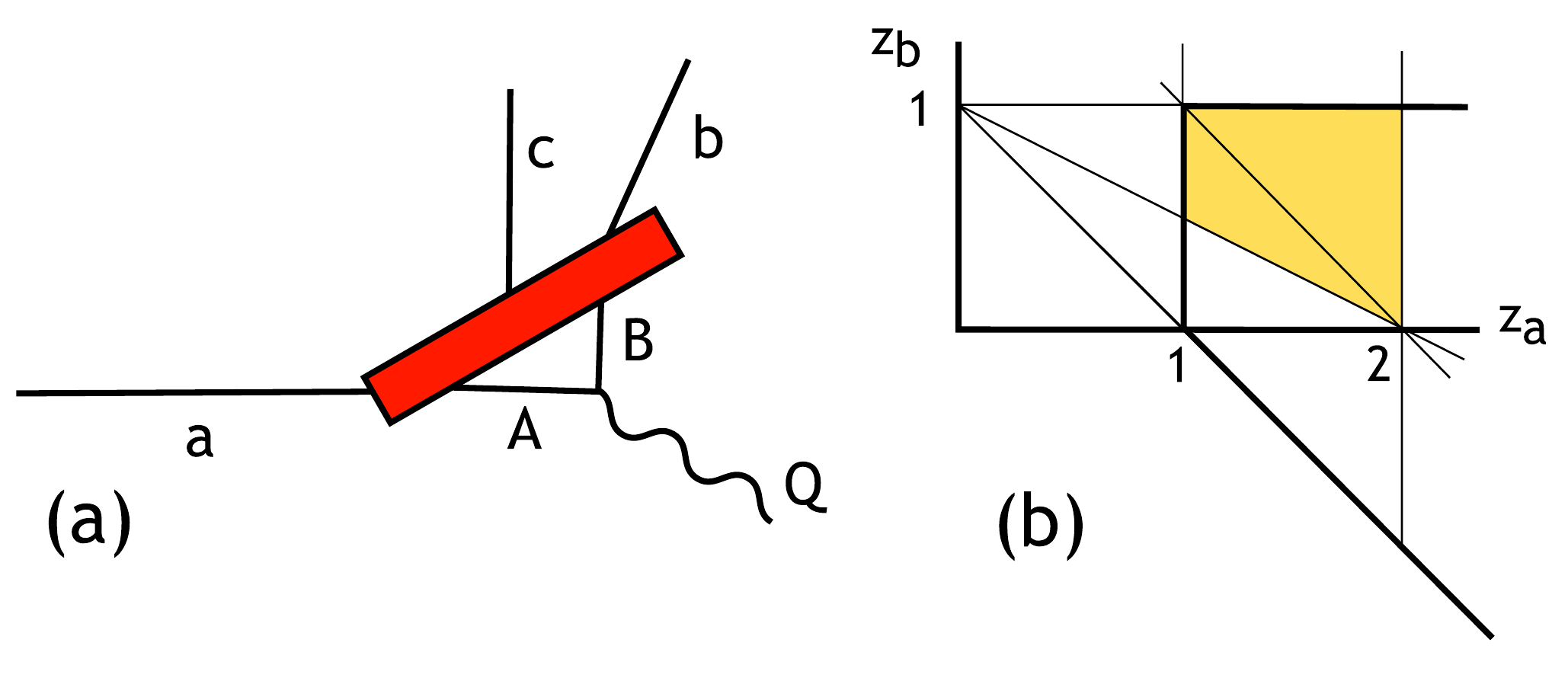}
\caption{(a) Kinematics of $2\to 3$ splitting in the initial-final (IF) case.
    (b) Phase space for $2\to 3$ splitting in the IF case.  The 
      eight regions corresponding to different orderings of 
    $|s_{ab}|$, $|s_{ac}|$, $s_{bc}$, $|Q^2|$ are shown. The region that should
       be well described by an antenna splitting $AB\to acb$ is shaded.}
\label{fig:IFkinematics}
\end{center}
\end{figure}

The decomposition of the amplitude is illustrated in 
Fig.~\ref{fig:IFkinematics}(a).  The kinematics can be described by 
variables $y_{ij}$ and $z_i$ obeying the relations \leqn{ydefs} to 
\leqn{yzmoreids}.  However, the vectors  $k_A$, $k_a$ now have
 negative timelike
component, and the vector $Q = k_A + k_B = k_a + k_b + k_c$ is spacelike,
$Q^2 = s_{AB} < 0$.  The phase space for this region covers the quadrilateral
shown in Fig.~\ref{fig:IFkinematics}(b).  The region of integration is
infinite, since $z_a$ can become very large, but the integral is cut off
at large $z_a$ by the parton distribution function.
The line $z_a > 1$, $z_b = 1$ 
corresponds to the region of initial state radiation, $c$ parallel to $a$.
 The line $z_a = 1$, $0 <z_b < 1$ 
corresponds to the region of final state radiation, $c$ parallel to $b$.
The line $z_a + z_b = 1$ corresponds to $b$ parallel to $a$, that is,
$b$ as initial state radiation from the primary $a$.   An 
       antenna shower should give an accurate description 
     of the dynamics in  the two regions   $|y_{ac}| < |y_{bc}| < 1$, 
 $|y_{bc}| < |y_{ac}| < |y_{ab}|$ that are shaded in the figure.  The new 
constraint $|y_{bc}| < 1$ is just $|s_{bc}| < |Q^2|$,
which is stronger than the constraint that this invariant is less
than $|s_{ab}|$. 

%%%%%%%%%%%%%%%%%%%%%%%%%%%%%%%%%%%%%%%%%%%%%%%%%%%%%%%%%%%%%%%%%%%%%%%%
%%%%%%%%%%%%%%%%%%%%%%%%%%%%%%%%%%%%%%%%%%%%%%%%%%%%%%%%%%%%%%%%%%%%%%%%%%%

To decompose \leqn{basicIF} into an appropriate form, we choose $p_A$ and 
$p_B$ and then change variables.  Let $p_A$ be chosen in the direction of
$p_a$, so that $p_a = z_a p_A$, $z_a > 1$.   
 Then $p_B = Q - p_A$.  We have
\beq
           p_a = x_a P \ , \quad  p_A = x_A P \ , \quad \mbox{so} \ 
                  \quad      x_a = z_a x_A \ , 
\eeq{xzforIF}
with $x_A$ having the definite value $x_A = -Q^2/2P\cdot Q$ associated with
scattering a massless particle from a local current.  For the reaction
$a Q \to bc$,  $s+t+u = Q^2$, so $t+u = Q^2 - s = Q^2 z_a$.  Then
\beq
t = Q^2 (1-z_b) =  \half Q^2 z_a (1-\cos\theta_*)
\eeq{zbzarel}
Using these formulae, we can change variables from 
 $(x_a, \cos\theta_*)$ to $(z_a, z_b)$.   The Jacobian of this transformation
is 
\beq
   J = { \del(x_a, \cos\theta_*)\over \del(z_a, z_b)} =  {2 x_A \over z_a}
\eeq{findJIF}
Thus,  
\beqa
\sigma (p X \to cb) &=& \int {dz_a\over z_a^2}
                  \int dz_b \int dx_A  x_A f(z_a x_A)  \ 
     \delta(x_A + Q^2/2P\cdot Q) \CR
       & & \hskip 1.6in \cdot  {1\over \Phi_{AX}} {1\over 8\pi}
           | \M(aX\to cb)|^2 \ .
\eeqa{exactIF}
This is an exact rewriting of \leqn{basicIF}.  Now apply the 
approximation \leqn{MdecompIF} and group terms to form
\beq
   \sigma(AX \to B) = {1\over \Phi_{AX}} 2\pi \delta(Q^2 + x_A 2P\cdot Q)
                             | \M(AX\to B)|^2 \ .
\eeq{sigmaIFform}
Then 
\beq
\sigma (p X \to cb) \approx \int {dz_a\over z_a^2}
      \int dz_b \int dx_A  f(z_a x_A)  
      \sigma(AX \to B)\cdot
  {\alpha_s N_c\over 4\pi}\, {\cal S}(z_a, z_c, z_b) \ .
\eeq{finalIF}

As an example, consider using this formula to describe initial-state
gluon radiation in deep inelastic scattering from a quark.  The total
gluon emission is given by the sum of the two spin-dependent
 splitting functions
in the fifth line of Table~\ref{tab:allN},
equal to
\beq
      \sum {\cal S} =  - {z_a^2 + z_b^2 \over y_{ac} y_{cb}}, 
\eeq{sumSIF}
The extra  minus sign comes from the sign of $s_{AB}$ in
 \leqn{Sdefintwo}.
In the region
of initial state radiation, $z_a = 1/w$, $z_b \approx 1$, 
$y_{ac} = -(1-z_b)$, $y_{bc} = (1 - 1/w)$.  Then, setting 
\beq
           \int dz_b {1\over 1-z_b} = \log{ Q^2\over \mu^2} \ , 
\eeq{getQsq}
we obtain
\beq
\sigma (p X \to cb) \approx \int dx_A \, \int {dw\over w}  
             f({x_A\over w}) \, 
      \sigma(AX \to B)\cdot
  {\alpha_s N_c\over 4\pi} {1 + w^2\over (1-w)}\log{Q^2\over\mu^2}  \ ,
\eeq{testfinalIF}
which is correct.

This is an appropriate point to discuss again the signs of
the expressions in Table~\ref{tab:allN}.  The antenna splitting functions
are probabilities; thus, they should be positive. However, we define the 
splitting functions in the IF and II regions as analytic continuations
of the values in the FF region, so their positivity must be checked 
explicitly. 

As we move from the 
FF region to the IF region with
$A$ and $a$ in the initial state, $y_{bc}$ becomes negative while
all other $y_{ab}$, $y_{ac}$, $z_a$, $z_b$ remain positive.  
The factor $z_c$ can be negative, but $z_c$ does not appear
in the Table.
With
the minus sign from $s_{AB}$ in  \leqn{Sdefintwo}, the denominator
of ${\cal S}(z_a,z_b,z_c)$ is positive, and so we need only check the 
numerator functions in given in the Table.
The numerator 
functions for $gg\to ggg$, $q\bar q \to q g \bar q$, and
 $qg\to qq\bar q$ remain positive, while
the numerator functions for $gg \to \bar q q g$ become negative.
In this last case, a fermion not present in the 2-parton system
is crossed from the final to the 
initial state, so we must supply an extra factor $(-1)$.  Then  all of
the expressions are positive, as required.  However, if we then 
cross from the region $z_b > 0 $ to the region $z_b < 0$,  one 
$qg\to qgg$ and one $qg \to q\bar q q$ amplitude changes sign.  
This sign change is unphysical; presumably, it is due to the 
simple method of reconstruction in \leqn{onetworeduce} and 
\leqn{onetworeducethree}.  We recommend setting these two 
amplitudes to zero for $z_b < 0$.
The region $z_b < 0$ is outside the shaded region in 
Fig.~\ref{fig:IFkinematics}(b) where we will generally use
the parton shower approximation, so most likely this
difficulty is not important in practice.

Similarly, for the FI region where $b$ and $B$ and taken to be in the
initial state, $y_{ac} < 0$.  Then
the numerators that go negative as we cross into the 
region are those in the $qg \to q\bar q q$ cases where a fermion is
crossed into the initial state. Now there are four amplitudes,  one
each in the $qg \to qgg$ cases and both of those in $q_-g_+\to 
q\bar q q$, that become negative when $z_a < 0$.  Again, we recommend
that these amplitudes be set to zero in this region of unphysical 
behavior.  

In the II region, both $y_{ac}$ and $y_{bc}$ are negative. The 
denominator of ${\cal S}(z_a,z_b,z_c)$ is positive.  The
numerator terms that are negative because of the sign changes
 are compensated by 
minus signs from crossing.  There are no unphysical sign changes.
The correct result is always obtained by taking the absolute
value of the numerator expression from Table~\ref{tab:allN}.

We now discuss the kinematics of the II case.
We begin from the formula for two protons of momentum
$P_A$, $P_B$ to produce a color-singlet system of momentum $Q$ plus a 
massless parton $c$,
\beq
  \sigma(pp \to cX) = \int dx_a \int dx_b f(x_a) f(x_b) \ {1\over 2s_{ab}}
     {1\over 16\pi} \int d\cos\theta_* {2p_*\over \sqrt{s_{ab}}}\,
                     |\M(ab\to cX)|^2 \ ,
\eeq{basicII}
where $\cos\theta_*$ and $p_*$ are the scattering angle and the momentum
in the $cX$ center of mass frame.

%%%%%%%%%%%%%%%%%%%%%%%%%%%%%%%%%%%%%%%%%%%%%%%%%%%%%%%%%%%%%%%%%%%%%%%%
\begin{figure}
\begin{center}
\includegraphics[height=2.0in]{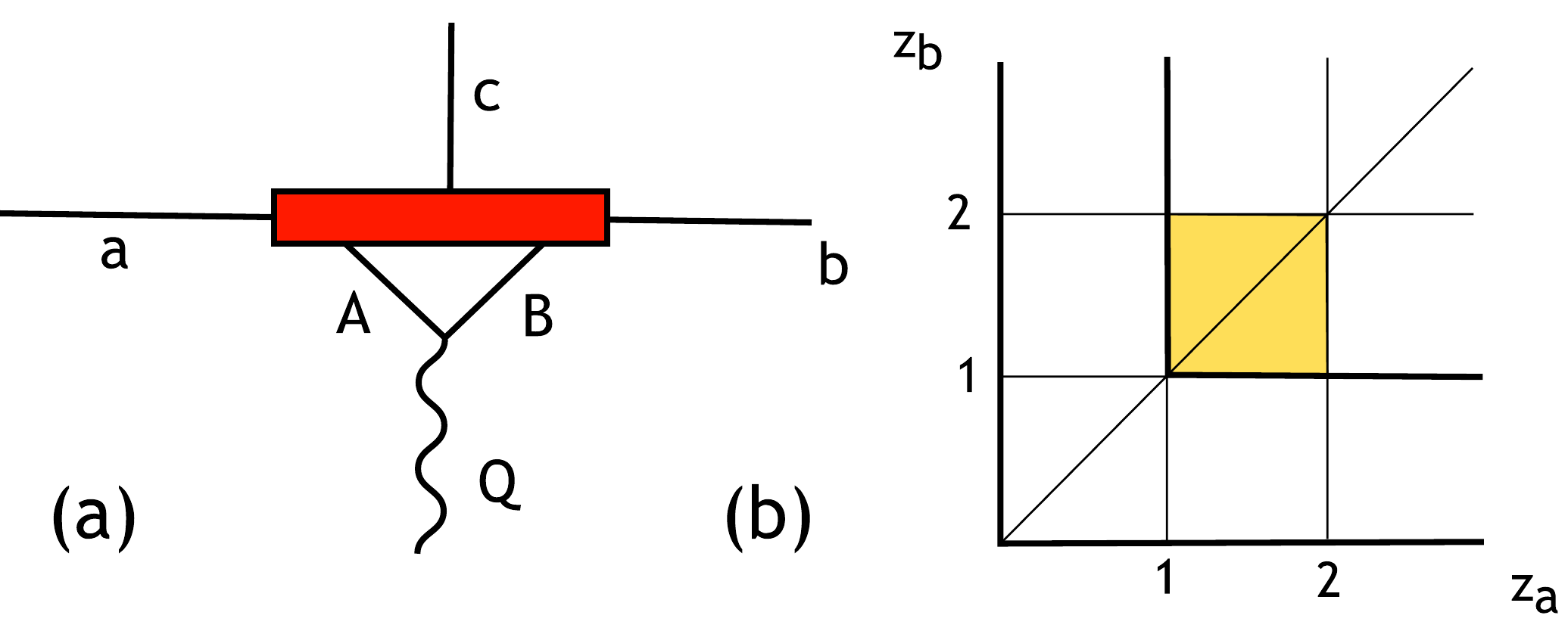}
\caption{(a) Kinematics of $2\to 3$ splitting in the initial state (II) case.
    (b) Phase space for $2\to 3$ splitting in the II case.  The 
      six regions corresponding to different orderings of 
      $|s_{ac}|$, $|s_{bc}|$, $|Q^2|$ are shown. The region that should
       be well described by an antenna splitting $AB\to acb$ is shaded.}
\label{fig:IIkinematics}
\end{center}
\end{figure}
%%%%%%%%%%%%%%%%%%%%%%%%%%%%%%%%%%%%%%%%%%%%%%%%%%%%%%%%%%%%%%%%%%%%%%%%%%%

The decomposition of the amplitude is illustrated in 
Fig.~\ref{fig:IIkinematics}(a).  The kinematics can again be described by 
variables $y_{ij}$ and $z_i$ obeying the relations \leqn{ydefs} to 
\leqn{yzmoreids}.  Now the vectors  $k_A$, $k_a$ $k_B$, $k_b$  have
 negative timelike
component, and the vector $Q = k_A + k_B = k_a + k_b + k_c$ is also 
negative timelike, with $Q^2 > 0$.
The phase space for this region covers the quadrant
shown in Fig.~\ref{fig:IFkinematics}(b), with $z_a, z_b > 1$. Again, 
the region of integration is
infinite, but the integral is cut off by the behavior of the parton 
distribution functions.
  The line $z_a > 1$, $z_b = 1$ 
corresponds to the region of initial state radiation with $c$ parallel to $a$.
The line $z_a = 1$, $z_b > 1$ 
corresponds to the region of initial state radiation with
$c$ parallel to $b$.
  An antenna shower should give an accurate description 
     of the dynamics in  the two regions   $|y_{ac}| < |y_{bc}| < 1$, 
 $|y_{bc}| < |y_{ac}| < 1$ that are shaded in the figure.  Again, the 
 limit 1 here corresponds to 
constraints $|s_{ac}|, |s_{bc}| < |Q^2|$,
which are stronger than the constraints that these two invariants are less
than $|s_{ab}|$. 

In the $ab\to cX$ process, the system $X$ must recoil with some nonzero
transverse momentum.  Thus, it is not possible to choose $k_A$ and $k_B$
to be parallel to $k_a$, $k_b$.  The invariants for the $ab\to cX$ scattering
process satisfy  $s + t + u = Q^2$.  Since $t = Q^2 (1-z_b)$,  
$u = Q^2(1- z_a)$, this means that $s = Q^2 (z_a + z_b -1)$. Alternatively,
$s = x_a x_b \cdot 2P_A \cdot P_B$.  We would like to choose the 
longitudinal fractions of $A$ and $B$,
$x_A$ and  $x_B$, to satisfy the relation
\beq
             x_A x_B \cdot 2P_A\cdot P_B = Q^2  \ .
\eeq{xAxBrel}
To make this possible, we must write
\beq
         x_a =  z_a x_A {\cal C}  \ ,  \qquad   x_b =  z_b x_B {\cal C} \ , 
\eeq{Cdefin}
with \cite{Daleoremark}
\beq
              {\cal C }^2 =   {z_a + z_b -1 \over z_a z_b}
\eeq{Cfind}
The function  ${\cal C}(z_a,z_b)$
approaches 1 when {\it either} $z_a$ or $z_b$ goes to 1;
that is ${\cal C} \approx 1$ in both collinear regions.

Also, $t + u = Q^2( 2 - z_a - z_b) = Q^2 z_c$, so 
\beq
      t = Q^2 (1 - z_b) =  \half Q^2 z_c (1 - \cos\theta_*)\ . 
\eeq{tforII}
We can now use \leqn{Cdefin} and \leqn{tforII} to 
change variables from $(x_a, x_b,\cos\theta_*)$  to $(x_A,z_a, z_b)$, 
holding $x_B$ fixed at the value $x_B = Q^2/x_A 2P_A\cdot P_B$.
 The Jacobian of this transformation
is 
\beq
   J = { \del(x_a, x_b, \cos\theta_*)\over \del(x_A, z_a, z_b)}  
= {2x_B\over z_c} = x_B {Q^2\over s_{ab}}{ \sqrt{s_{ab}}\over p_*} \ .
\eeq{findJII}
Then  
\beqa
\sigma (pp \to cX) &=& \int {dz_a\over z_a^2} {dz_b\over z_b^2}{1\over 
            {\cal C}^4}
                  \ \int dx_A dx_B\, f(z_a x_A {\cal C})  f(z_b x_B {\cal C})
           \ x_B \delta(x_B - Q^2/x_A 2P_A\cdot P_B)  \CR
       & & \hskip 1.6in \cdot  {1\over s_{AB}} {1\over 8\pi}
           | \M(aX\to cb)|^2 \ .
\eeqa{exactII}
This is an exact rewriting of \leqn{basicII}.  Now apply the 
approximation analogous to \leqn{MdecompFF} or \leqn{MdecompIF} 
and group terms to form
\beq
   \sigma(AB \to X) = {1\over 2 s_{AB}} 2\pi 
             \delta(Q^2 - x_A x_B 2P_A\cdot P_B)
                             | \M(AX\to B)|^2 \ .
\eeq{sigmaIIform}
This gives, finally,
\beq
\sigma (pp \to cX) \approx \int {dz_a\over z_a^2} {dz_b\over z_b^2}{1\over 
            {\cal C}^4}
      \ \int dx_A dx_B\,  f(z_a x_A {\cal C})  f(z_b x_B {\cal C})  
      \sigma(AB \to X)\cdot
  {\alpha_s N_c\over 4\pi}\, {\cal S}(z_a, z_c, z_b) \ .
\eeq{finalII}

To test this formula, consider the case of $q \bar q$ annihilation with 
the emission of a gluon collinear with the quark $a$.  The sum of 
 spin-dependent splitting functions for this case is again \leqn{sumSIF}.
In the collinear region of interest, $z_a = 1/w$, $z_b\approx 1$. 
Repeating the step that led to \leqn{testfinalIF}, we find 
\beq
\sigma (p p \to cX) \approx \int dx_A dx_B\, \int {dw\over w}  
             f({x_A\over w})  f(x_B) \,
      \sigma(AB \to X)\cdot
  {\alpha_s N_c\over 4\pi} {1 + w^2\over (1-w)}\log{Q^2\over\mu^2}  \ ,
\eeq{testfinalII}
which is the correct limit.

\section{Comparison to previous results}

In the Introduction, we made reference to a number of previous 
definitions of the antenna splitting 
functions.
We noted that these definitions agree, as they must, in the singular
soft and collinear limits.  However, these prescriptions differ
widely away from the boundaries of phase space.  In this section, we 
will compare our prescription to those of ARIADNE~\cite{colordipoles,ARIADNE}
 and Gehrmann-De Ridder, \etal~\cite{Gehrmann}.  

We will make this comparison over the natural
phase space discussed in the previous section--the entire $(z_a,z_b)$ plane
above the line $z_a + z_b = 1$.  In order to describe antenna showers
for initial- as well as final-state emissions, the splitting functions 
should extend into the region $z_a, z_b > 1$. Depending on the details of how
the shower is constructed, their use might be restricted to a polygon around
$z_a = z_b = 1$, or the expressions might be used for arbitrarily large
values of $z_a$ and $z_b$.

We note again that the IF regions include the lines $z_a = 0$ and 
$z_b = 0$.  Expressions
for the splitting functions that are well-behaved near $z_a = z_b = 1$ 
can possibly have a singularity on this line, though such a singularity in 
the middle of the phase space would be unphysical.  We used this criterion
in Section 4 to exclude factors of $1/z_a$ and $1/z_b$ from appearing 
in \leqn{onetworeducethree}.  The antenna functions of Duhr and 
Maltoni~\cite{Duhr} are typically singular along this line and so cannot
be used in parton shower models in all regions.

The ARIADNE and Gehrmann-De Ridder
antenna functions give expressions summed over final
polarizations.   
To compare our splitting functions to these, we must sum over a row
in Table \ref{tab:allN}.   Our summed expressions are independent of the 
initial polarization in the soft and collinear limits, but they depend on 
the polarizations of $A$ and $B$ in the interior of the $(z_a,z_b)$ space.
The comparison to our expressions thus also reveals where this dependence
on polarization is an important effect. 

The first antenna splitting functions were put forward
 by the ARIADNE group \cite{colordipoles}.  Their 
approach started from the spin-averaged cross section 
for the simple splitting process $q\bar{q}\rightarrow q 
g\bar{q}$ in $\ee$ annihilation.  They then guessed the expressions
for the $qg\rightarrow qgg$ and $gg\rightarrow ggg$ splittings, 
so that these would have a similar form to the $q\bar{q}\rightarrow q 
g\bar{q}$ case,
\beq
           {\cal S} =   { z_a^{n_a} + z_b^{n_b} \over y_{ac} y_{bc}} \ ,
\eeq{ARIADNESproposal}
where $n_a, n_b$ = 2 for emission from a quark and 3 for emission from a
gluon.

Our philosophy, explained in Section 2, is that each individual antenna
should reproduce the collinear limit predicted by QCD.  These expressions
are symmetric under interchange of identical particles, while 
\leqn{ARIADNESproposal} does not have this property, so we would obtain
the complete splitting function by symmetrizing \leqn{ARIADNESproposal}.
This gives 
\beqa
q\bar{q}\:\:{\rm antenna\!:}&\:\:{\cal S} 
&= {z_a^2+z_b^2 \over y_{ac}y_{bc}} \ ,\CR
gg\:\:{\rm antenna\!:}&\:\:{\cal S} 
&= {z_a^3+z_b^3\over y_{ac} y_{bc}} + {z_a^3+z_c^3\over y_{ab} y_{bc}}
           + {z_b^3+z_c^3\over y_{ab}y_{ac}} \ ,\CR
qg\:\:{\rm antenna\!:}&\:\:{\cal S} 
&= {z_a^2+z_b^3\over y_{ac}y_{bc}} + {z_a^2+z_c^3\over y_{ab} y_{bc}}\ .
\eeqa{ARIADNEant}
The summed terms are each positive in the FF kinematic region.  To obtain
the ARIADNE splitting functions 
in the other regions, we analytically continue these 
formulae into the regions where $z_a$ or $z_b$ is greater than 1.

The analytic continuation of the ARIADNE and, below, the 
Gehrmann-de Ritter results brings in the issue of the positivity
of these expressions, similar to the positivity issue for our 
splitting functions discussed in Section 6. 
For the ARIADNE and Gehrmann-De Ridder antenna functions, the expressions
given are summed over spins, and the individual pieces are not independent
of one another.  So, if they become negative, that is a problem for
the complete, spin-summed, expression.  For the Gehrmann-de Ridder
functions, it can be seen that this happens
only the regions $z_a < 0$ and $z_b < 0$, so this is not a serious
problem.  However, the ARIADNE function involve $z_c^3$, which is
negative in the whole region $z_a + z_b > 2$.  This problem cannot
be resolved by replacing $z_c$ with $|z_c|$, since this leads to 
expressions that do not agree with the Altarelli-Parisi factorization
along the lines separating the IF regions from the II region. Fortunately,
the ARIADNE functions do not become actually become negative until 
$z_a$ or $z_b$ becomes very large ($z_a$ or $z_b \sim 12$).  
However, the idea that
the ARIADNE functions are sums of positive and negative terms in the 
initial-state regions goes against the intuition used to propose these
expressions.

\begin{figure}
\begin{center}
{
    \includegraphics[width=7.2cm]{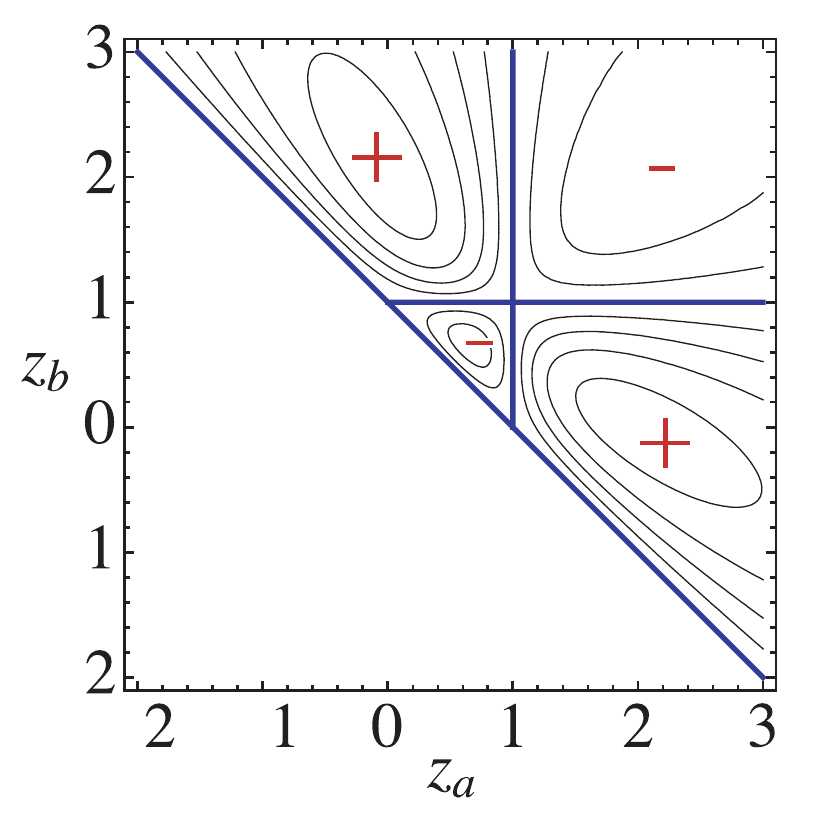}
}
\hspace{0cm}
{
    \includegraphics[width=7.2cm]{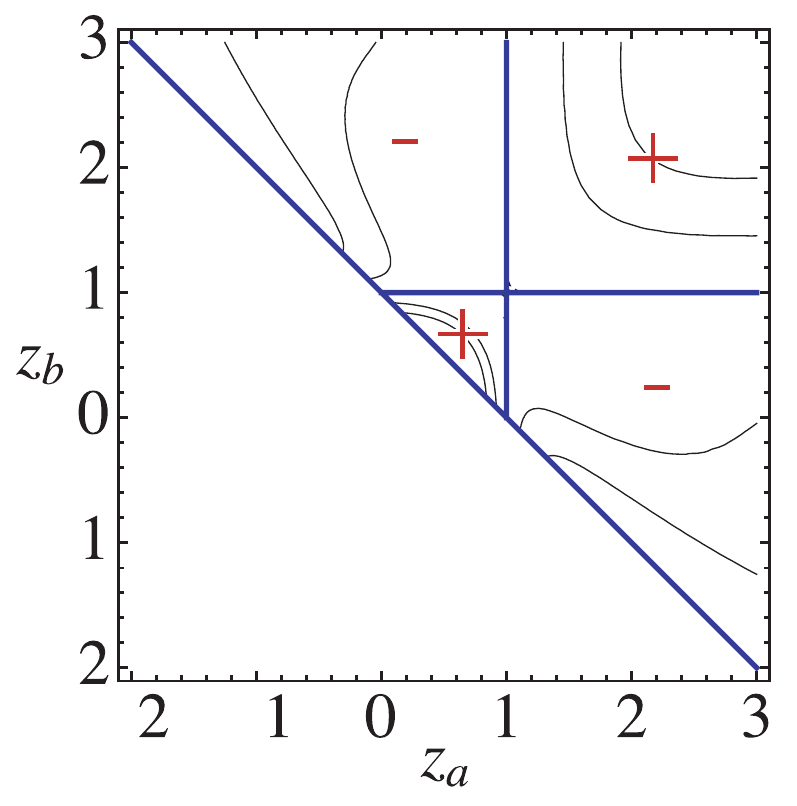}
}
\caption{Visualization of the ratio of the ARIADNE antenna 
function to our antenna functions for the 
processes $gg\rightarrow ggg$.  The figures on the left and 
right are the comparison of the ARIADNE 
antenna function to our spin-summed antenna functions from 
row 1 and row 2 in Table \ref
{tab:allN}, respectively.  The boundaries of phase space for
the different kinematic regions are marked in blue.  The contours 
are plotted at ratios of 1.2, 1.5, 2.0, 3.0, and 5.0, with $+$ 
indicating a region in which the ratio is greater than 1.}
\label{fig:ARIADNEggg}
\end{center}
\end{figure}

\begin{figure}
\begin{center}
    \includegraphics[width=7.2cm]{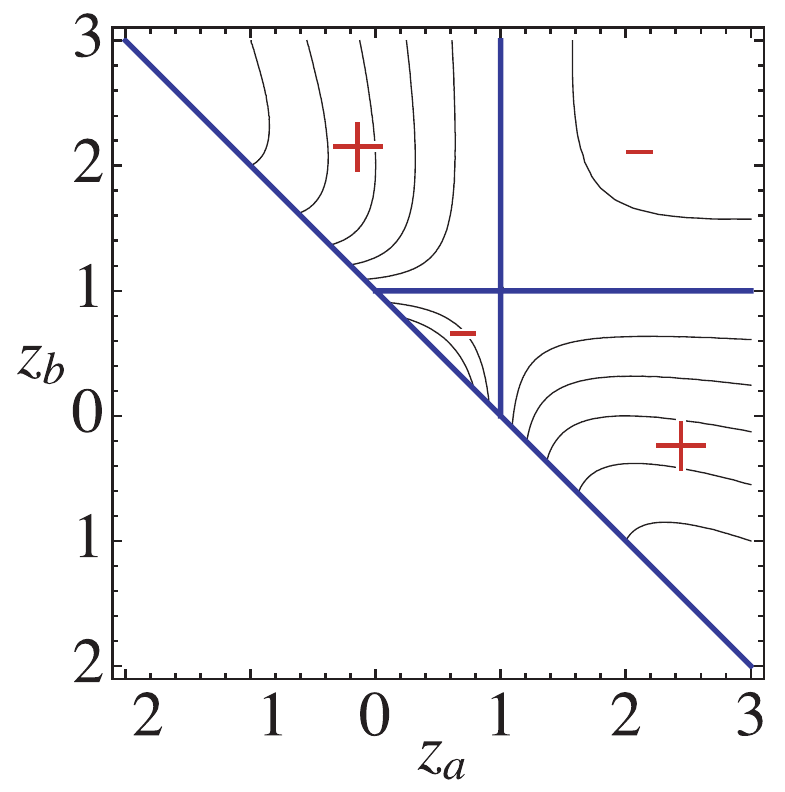}
\caption{Visualization of the ratio of the ARIADNE antenna
 function to our antenna function for the 
process $q_-\bar{q}_-\rightarrow qg\bar{q}$.  Our antenna
 function for the process $q_-\bar{q}_+\rightarrow 
qg\bar{q}$ coincides with the ARIADNE result and so is not
 included.
 The notation is as in Fig.~\ref{fig:ARIADNEggg}.}
\label{fig:ARIADNE6}
\end{center}
\end{figure}

\begin{figure}
\begin{center}
{
    \includegraphics[width=7.2cm]{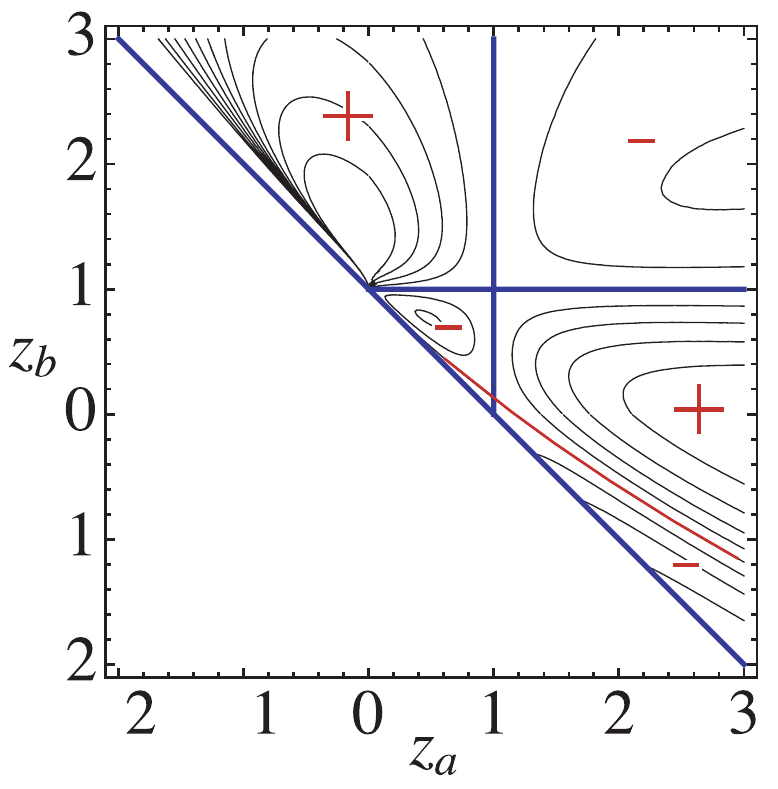}
}
\hspace{0cm}
{
    \includegraphics[width=7.2cm]{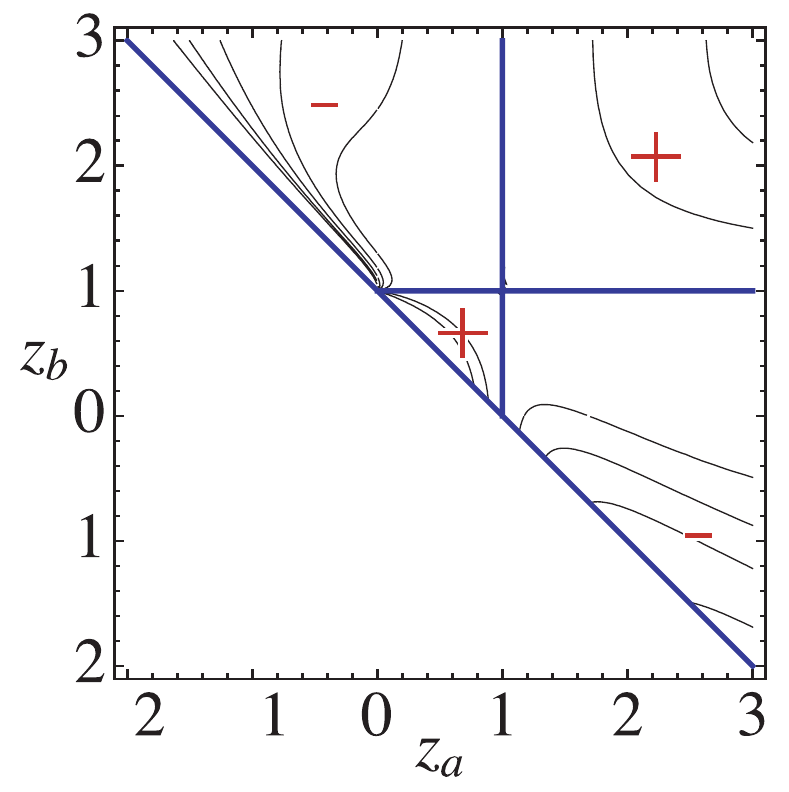}
}
\caption{Visualization of the ratio of the ARIADNE antenna
 function to our antenna functions for the 
processes $qg\rightarrow qgg$.  The figures on the left and
 right are the comparison of the ARIADNE 
antenna function to our spin-summed antenna functions from
 row 7 and row 8 in Table \ref{tab:allN}, respectively. 
The notation is as in Fig.~\ref{fig:ARIADNEggg}.}
\label{fig:ARIADNEqgg}
\end{center}
\end{figure}

We are now in a position to compare the ARIADNE function to our proposal.
For the $q\bar q$ antenna, the expression above coincides with the 
sum of row 5 of Table 
\ref{tab:allN}.   For the $gg$ and $gq$ cases, the ratio of the above
ARIADNE 
functions to those defined in Table \ref{tab:allN} are illustrated in 
Figs.~\ref{fig:ARIADNEggg}, \ref{fig:ARIADNE6}, 
and \ref{fig:ARIADNEqgg}.  The notation in the figures is the following:
Each figure represents the ratio of the ARIADNE splitting function to 
our results 
for a specific initial set of polarized partons,  summed over final
state polarizations.  The ratio goes to 1 on 
the lines $z_a = 1$ and $z_b = 1$, which correspond to the collinear
limits.  Away from these lines, the contours on which the ratios are
1.2, 1.5, 2.0, 3.0, and 5.0 (toward the $+$ symbol), 
and the inverses of these numbers (toward the $-$ symbol) are shown.  
The $qg$ antenna function are asymmetric between partons $a$ and $b$.
The IF region in the lower right is that in which the quark is in the
intial state and the gluon is in the final state.  The IF region
in the upper left is that in which the gluon is in the initial state and
the quark remains in the final state.

%%%%%%%%%%%%%%%%%%%%%%%%%%%%%%%%%%%%%%%%%%%%%%%%%%%%%%
%%%%%%%%%%%%%%%%%%%%%%%%%%%%%%%%%%%%%%%%%%%%%%%%%%%%%%%%

%%%%%%%%%%%%%%%%%%%%%%%%%%%%%%%%%%%%%%%%%%%%%%%%%%%%%%
%%%%%%%%%%%%%%%%%%%%%%%%%%%%%%%%%%%%%%%%%%%%%%%%%%%%%%%%

%%%%%%%%%%%%%%%%%%%%%%%%%%%%%%%%%%%%%%%%%%%%%%%%%%%%%%
%%%%%%%%%%%%%%%%%%%%%%%%%%%%%%%%%%%%%%%%%%%%%%%%%%%%%%%%

The ARIADNE authors gave a different interpretation to the formulae
\leqn{ARIADNEant}.  They took the philosophy that the collinear
limit need not result from a single antenna but rather should be
the result of summing over the possible antennae that would lead to 
a specific final state.  A three gluon final state could result from 
any pair of the gluons radiating the third and so should be the sum 
of three antennae.  Then the second line of 
\leqn{ARIADNEant} would be interpretated as the sum over these three antennae.
This is a reasonable point of view for the FF kinematics considered in
\cite{colordipoles}.  However, in the  IF and II regions, at least one
of the $z_i$ will be negative and so some of the 
terms in the last two lines of \leqn{ARIADNEant} will become negative.
Such terms cannot be interpreted as independent radiators, each emittting
a gluon with positive probability.  It is tempting to revise the formula
in \leqn{ARIADNEant} by taking the absolute
values of the negative terms.   However, one can readily check that no such
prescription gives the correct Altarelli-Parisi limit along the lines
$z_a = 1$ and $z_b =1$ at the boundaries of the IF and II regions.
Thus, we believe, the ARIADNE formulae can be used in the IF and II regions
only by using the formulae \leqn{ARIADNEant} as written and accepting 
that some negative signs will appear~\cite{thankJan}.

\begin{figure}
\begin{center}
{
    \includegraphics[width=7.2cm]{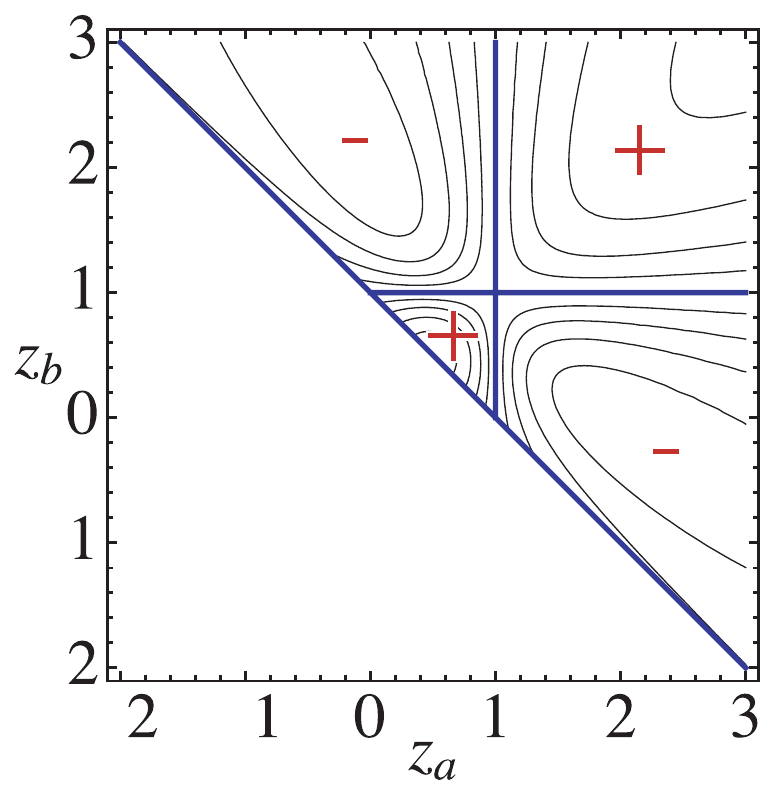}
}
\caption{Visualization of the ratio of the Gehrmann-De Ridder antenna
 function to our antenna function for the 
process  $g_-g_+\rightarrow ggg$.  The antenna
 function for the process $g_+g_+\rightarrow 
ggg$ coincides with the Gehrmann-De Ridder result and so is not
 included.
 The notation is as in Fig.~\ref{fig:ARIADNEggg}.}
\label{fig:Gehrmann2}
\end{center}
\end{figure}

Gehrmann-De Ridder, Gehrmann and Glover \cite{Gehrmann} studied $2\to 3$
splitting from Feynman diagrams to develop an antenna subtraction program 
for NNLO calculations.  In doing 
so, they were able to extract 
unpolarized antenna functions for the processes 
$gg\rightarrow ggg$, $qg\rightarrow 
qgg$ and $qg\rightarrow q\bar{q}q$.  To calculate the gluon-gluon 
antenna function, they used the effective 
Higgs coupling to gluons 
\beq
{\cal L} = - {\lambda \over 4} h F^{\mu\nu}F_{\mu\nu}.
\eeq{GehrmannHgg}
This is essentially the same procedure that we used in Section 3, and it 
yields the same result as the sum of row 1 in 
Table \ref{tab:allN}.  In our language, their antenna function 
for the gluon-gluon dipole is~\cite{factorthird}
\beq
{\cal S} = {y_{ac}^2+y_{bc}^2+y_{ab}^2+y_{ac}^2y_{bc}^2
+y_{ab}^2y_{bc}^2+y_{ab}^2y_{ac}^2 \over y_{ab}y_{ac}y_{bc}} + 4\ .
\eeq{Gehrmannggant}
  The comparison of this antenna function to the sum of row 2 of 
Table \ref{tab:allN} is illustrated in Fig.~\ref{fig:Gehrmann2}.

This splitting function for $gg \rightarrow ggg$ is, however, 
not precisely the form of the
 splitting function that is used in the VINCIA parton shower~\cite{VINCIA}.  
They use the `global' form of the Gehrmann-De Ridder antenna function, 
 which in our language is
\beq
{\cal S} = \half \left[ \frac{2 y_{ab}^2+y_{ab}^2y_{ac}^2
 +y_{ab}^2y_{bc}^2}{y_{ab}y_{ac}y_{bc}}+\frac{8}{3} \right].
\eeq{VINCIAimp}
To implement this antenna function, a similar procedure is used as with the 
ARIADNE antenna 
functions.  That is, emissions from overlapping 
antenna are summed.  When the three antennae contributing 
to $gg\to ggg$ are summed together, one recovers the result 
\leqn{Gehrmannggant}.  This prescription
works well in the FF kinematics. However, as
in the ARIADNE case, it might require negative contributions in
 splitting functions for 
some antennae in the IF and II kinematics.

\begin{figure}
\begin{center}
{
    \includegraphics[width=7.2cm]{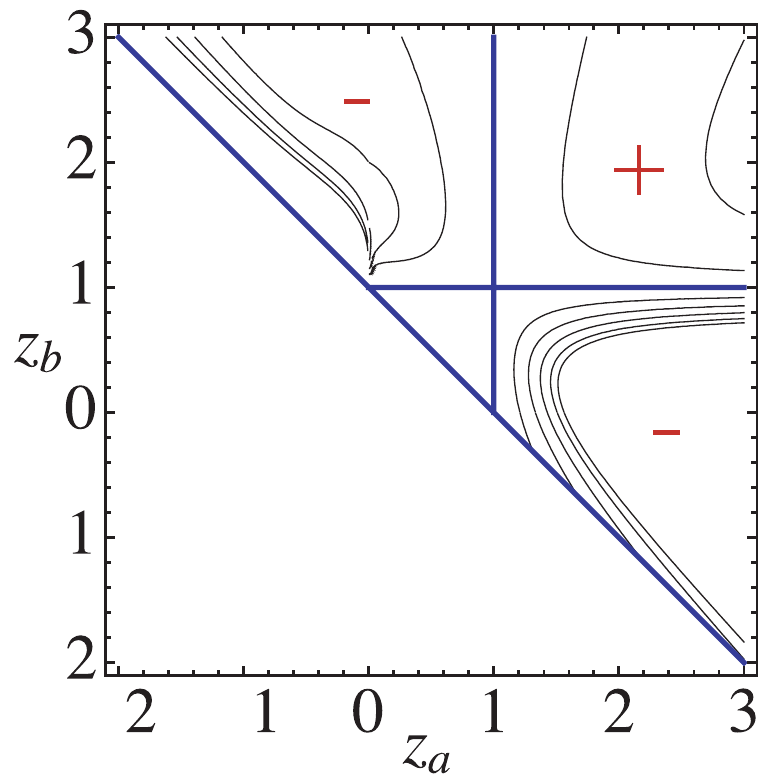}
}
\hspace{0cm}
{
    \includegraphics[width=7.2cm]{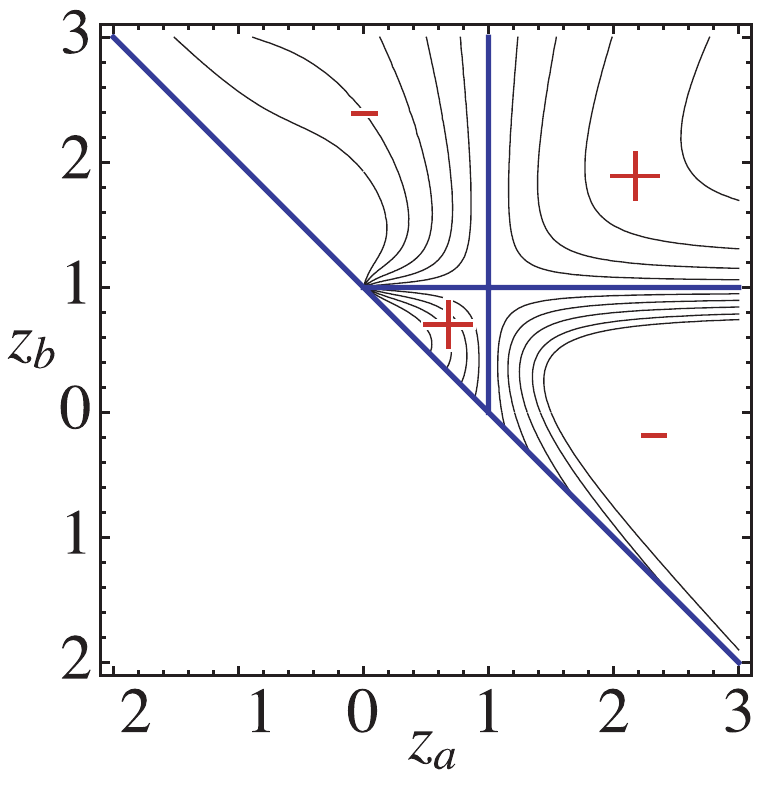}
}
\caption{Visualization of the ratio of the Gehrmann-De Ridder antenna
 functions to our antenna functions for the 
processes   $qg\rightarrow qgg$. The figures on the left and right are 
the comparison of the 
Gehrmann-De Ridder 
antenna function to our spin-summed antenna functions from 
row 7 and row 8 in Table \ref{tab:allN}, respectively. 
 The notation is as in Fig.~\ref{fig:ARIADNEggg}.}
\label{fig:Gehrmannqgg}
\end{center}
\end{figure}

\begin{figure}
\begin{center}
{
    \includegraphics[width=7.2cm]{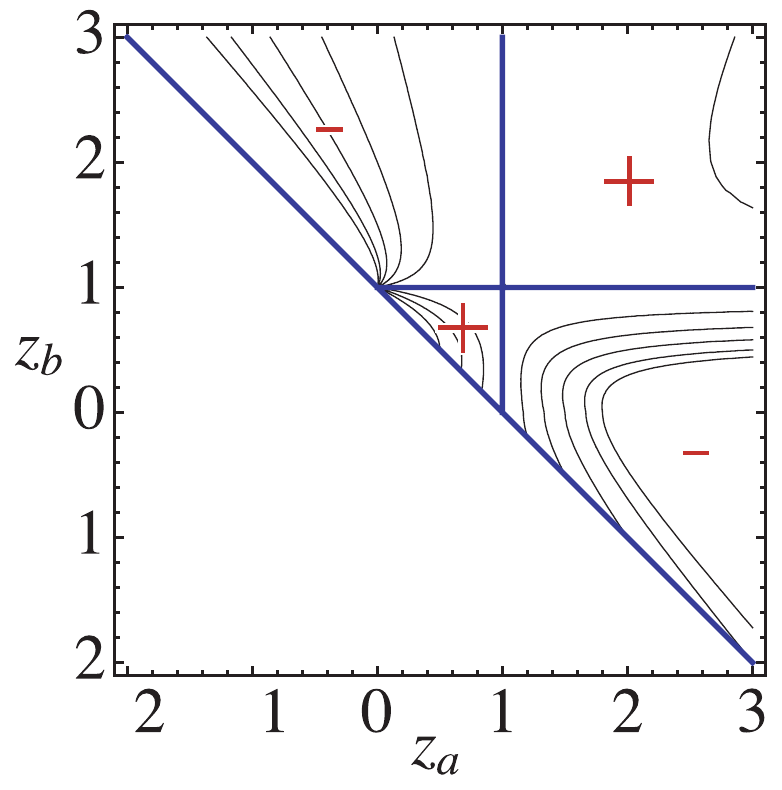}
}
\hspace{0cm}
{
    \includegraphics[width=7.2cm]{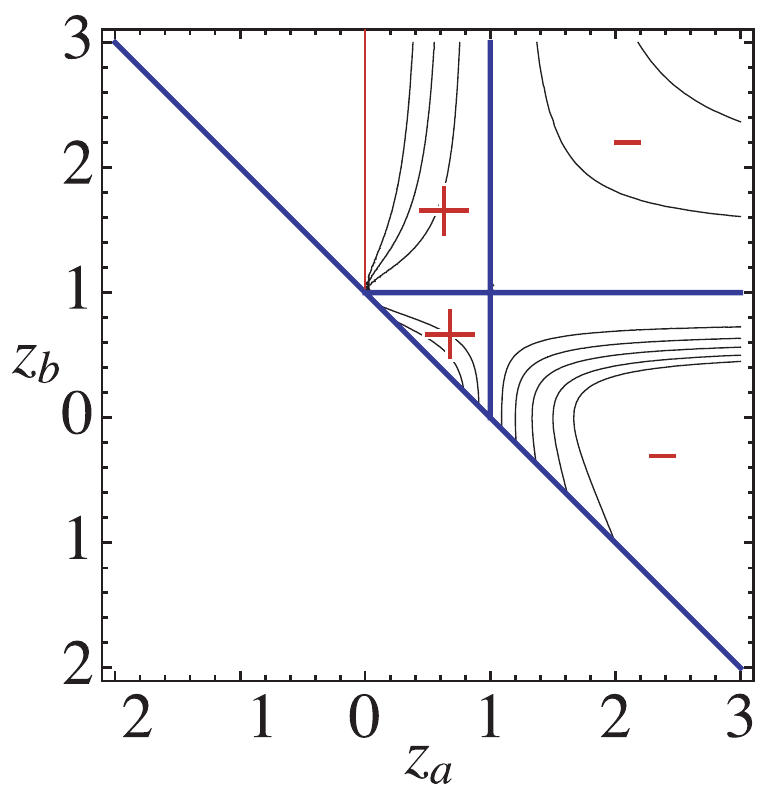}
}
\caption{Visualization of the ratio of the Gehrmann-De Ridder antenna
 functions to our antenna functions for the 
processes  $qg\rightarrow qq\bar{q}$. The figures on the left and right are 
the comparison of the 
Gehrmann-De Ridder
antenna function to our spin-summed antenna functions from 
row 9 and row 10 in Table \ref{tab:allN}, respectively. 
 The notation is as in Fig.~\ref{fig:ARIADNEggg}.}
\label{fig:Gehrmannqqq}
\end{center}
\end{figure}

To construct the antenna functions involving quarks, Gerhmann-De Ridder,
\etal, calculated 
the decay of a neutralino $\chi$ to a 
gluon and a gluino $\psi$ through the effective operator
\beq
{\cal L}=i\eta \bar{\psi} \sigma^{\mu\nu}\chi F_{\mu\nu}+ \hc
\eeq{Gehrmannqg}
In principle, our results should agree for case of a spin $\half$ 
initial state.  However, 
our choices \leqn{onetworeduce} and \leqn{onetworeducethree}
for handling ambiguous momentum products, produce some differences.
  In our language, their antenna functions involving quarks are
\beqa
qg\rightarrow qgg:&\:\:{\cal S} &= {2y_{ab}^2+2y_{ac}^2
+y_{ab}y_{bc}^2+y_{ac}y_{bc}^2+2y_{ac}^2y_{ab}^2 
\over y_{ab}y_{ac}y_{bc}}+2+2y_{ac}+2y_{ab} \ ,\CR
qg\rightarrow q\bar{q}q:&\:\:{\cal S} 
&= {(y_{ac}+y_{ab})^2y_{ac}y_{ab}-2y_{ac}^2y_{ab}^2\over y_{ab}y_{ac}y_{bc}}
+y_{ab}+y_{ac} \ .
\eeqa{Gehrmannqgant}
The comparison to our antenna functions is illustrated in 
Figs.~\ref{fig:Gehrmannqgg} and \ref{fig:Gehrmannqqq}.  For
$qg\to qgg$, our result for the spin $\half$ case is indeed very close to 
the above expression in the FF region.  
For $qg\to q\bar q q$, our prescription
\leqn{onetworeducethree}  gives us an extra factor of $z_a$ near $z_a = 0$.

%%%%%%%%%%%%%%%%%%%%%%%%%%%%%%%%%%%%%%%%%%%%%%%%%%%%%%
%%%%%%%%%%%%%%%%%%%%%%%%%%%%%%%%%%%%%%%%%%%%%%%%%%%%%%%%

%%%%%%%%%%%%%%%%%%%%%%%%%%%%%%%%%%%%%%%%%%%%%%%%%%%%%%
%%%%%%%%%%%%%%%%%%%%%%%%%%%%%%%%%%%%%%%%%%%%%%%%%%%%%%%%
%%%%%%%%%%%%%%%%%%%%%%%%%%%%%%%%%%%%%%%%%%%%%%%%%%%%%%
%%%%%%%%%%%%%%%%%%%%%%%%%%%%%%%%%%%%%%%%%%%%%%%%%%%%%%%%

In summary, we have shown that the antenna splitting functions represented
by \leqn{Sproposal} and Table~\ref{tab:allN} give a physically sensible
prescription for the construction of antenna showers.  These splitting
functions can  be used with the formulae \leqn{finalFF}, 
\leqn{finalIF}, \leqn{finalII} to generate antenna splittings in all
three relevant kinematic regions.  We hope that this formalism will 
provide a firm foundation for the construction of new parton showers
based on the antenna concept.

\Acknowledgements

We thank Darren Forde, Tanju Gleisberg, Peter Skands,
 and Jan Winter for instructive discussions and  Claude Duhr and 
Fabio Maltoni for a useful correspondence.
This work was aided by our participation in the Northwest Terascale
Workshop on Parton Showers and Event Structure at the LHC at the
University of Oregon. We thank
the participants and, especially, the organizer, Davison Soper.
  The work was supported
by the US Department of Energy under contract DE--AC02--76SF00515.


\begin{thebibliography}{99}
\bibitem{correctoneoverN}
 In writing a parton shower for $\ee\to q\bar q$, one might
   pick up an obvious $1/N_c$ correction by writing the prefactor
   as ${8\over 3}$ instead of $3$ for the first gluon emission
      from the quark line.
            In higher orders, the analogous treatment of radiation
          from quark lines
      is not so innocent and can potentially lead to incorrect
               accounting of color coherence.
\bibitem{factorthird}
This expression actually differs from the one given in \cite{Gehrmann} 
by a factor of $1/3$, which 
comes from allowing any gluon to become collinear with any other gluon.
Since we have identified 
the radiated gluon, we remove this factor.
\end{thebibliography}
\end{document}